\documentclass[runningheads]{llncs}
\usepackage[T1]{fontenc}
\usepackage{amsmath}
\usepackage[ruled,vlined]{algorithm2e}
\usepackage{booktabs}
\usepackage{longtable}
\usepackage{graphicx}
\usepackage{float}
\usepackage[utf8]{inputenc}
\usepackage{algorithmic}
\usepackage{hyperref}
\usepackage{makecell}
\usepackage{listings}
\begin{document}

\title{A Multi-tiered Solution for Personalized Baggage Item Recommendations using FastText and Association Rule Mining}

\author{Mudavath Ravi\inst{1} \and Atul Negi\inst{1}\thanks{Corresponding Author: Atul Negi (\email{atul.negi@uohyd.ac.in})}}

\institute{School of Computer and Information Sciences, University of Hyderabad, Hyderabad 500046, Telangana, India}

\maketitle

\begin{abstract}
This paper introduces an intelligent baggage item recommendation system to optimize packing for air travelers by providing tailored suggestions based on specific travel needs and destinations. Using FastText word embeddings and Association Rule Mining (ARM), the system ensures efficient luggage space utilization, compliance with weight limits, and an enhanced travel experience. The methodology comprises four phases: (1) data collection and preprocessing with pre-trained FastText embeddings for text representation and similarity scoring (2) a content-based recommendation system enriched by user search history (3) application of ARM to user interactions to uncover meaningful item associations and (4) integration of FastText and ARM for accurate, personalized recommendations. Performance is evaluated using metrics such as coverage, support, confidence, lift, leverage, and conviction. Results demonstrate the system's effectiveness in providing relevant suggestions, improving customer satisfaction, and simplifying the packing process. These insights advance personalized recommendations, targeted marketing, and product optimization in air travel and beyond.

\end{abstract}

\section{Introduction}
\par{Stress has become an inevitable companion due to the modern lifestyle. To reduce stress, the allure of travel offers an escape.  Most travelers plan their trip to fulfill their travel aspirations and have good travel memories \cite{lee2017ontology}. Airtravel is now common place for many segments of society including the lesser developed nations. Increased use of automated systems and direct advice to users is very important. The web is a very valuable source of travel information replacing traditional sources of knowledge. Travel information and suggestions for flights, hotels, vacation packages, etc., are available on specialized websites like WikiVoyage and Frommers, Expedia or Skyscanner, TripAdvisor, etc. Even though all of these sites are excellent sources of information, they rarely offer specific recommendations about the items to carry in the baggage so that a traveler going for business or pleasure may complete the trip without baggage-related hassles. The insights gained from this study extend beyond luggage packing optimization, potentially influencing targeted marketing strategies and product optimization within the travel industry Only in air transportation severe baggage problems or baggage security problems are encountered hence the focus of this paper is on air travel recommendations.}
\par{In recent times air travel baggage is a significant concern to passengers and also the air transportation system like airports, airlines, and regulatory authorities. The goal of travel recommendation systems is to match the user's needs with the characteristics of travel and leisure resources and attractions. It makes sense to construct recommendation systems for customized travel packages. This is a novel approach. As of now, to the best of our knowledge, there is no baggage advice recommendation system present. Our aim for this purpose is that to improve overall packing efficiency of airports where passengers are well prepared even before undergoing lengthy check-in process.}
\par{Recommendation systems, or recommender systems, are software engines designed to suggest products to users based on their past preferences, engagement, and interactions. These systems utilize filtering techniques to predict user preferences, interests, or needs, providing a personalized experience. In the context of the air travel industry, an Air Travel Baggage Recommender System (ATRS) serves as an information-filtering engine tailored for travel-related recommendations \cite{sarkar2023tourism}. This includes personalized flight options, travel packages, seat preferences, and destination suggestions, considering user preferences, travel history, and relevant criteria.}
\par{Similar to recommendation engines, an ATRS aids travelers in finding suitable flight options by analyzing past travel patterns, customer feedback, and other data, streamlining the booking process and enhancing user satisfaction. The typical recommendation system workflow involves data collection, preprocessing, feature extraction, model training, recommendation generation, evaluation, feedback collection, system updates, alpha-beta testing, and monitoring and maintenance post-deployment. This study focuses on the specific nuances of an ATRS, exploring how it can be customized to optimize recommendations in the air travel sector.}
\subsection{About ATRS}
The concept of integrating baggage-related advice into an Air Travel Baggage Recommender System (ATRS) represents a novel proposition with potential benefits for the air travel industry. In this context, travelers are treated as users, and items are identified as baggage. The proposed recommender system aims to calculate the flexibility that a traveler desires in interacting with baggage items, utilizing this information to generate tailored recommendations for the most suitable subset of baggage. By incorporating baggage-related advice into the ATRS, the system could significantly enhance the overall travel experience for users, offering personalized suggestions that align with their preferences and needs \cite{pelanek2024personalized}. This innovative approach has the potential to revolutionize how travelers engage with and select baggage, providing a valuable and unique feature in the realm of air travel recommendation systems.
\subsection{Motivation of Baggage Module in ATRS}
\begin{figure}[H]
\centering
\fbox{\includegraphics[width=0.9\linewidth]{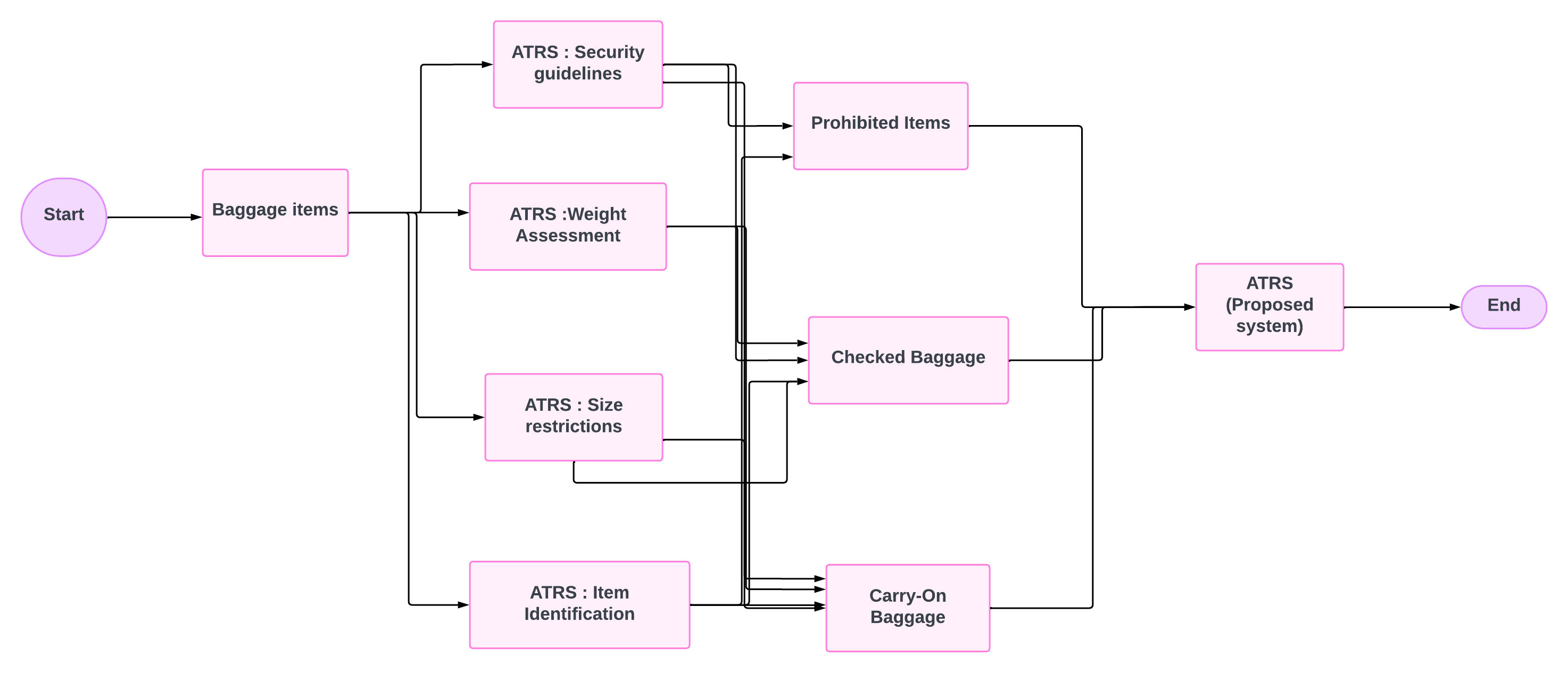}} \caption{ATRS Concept}
\label{fig:ag}
\end{figure}

\begin{figure}[H]
\centering
\fbox{\includegraphics[width=0.9\linewidth]{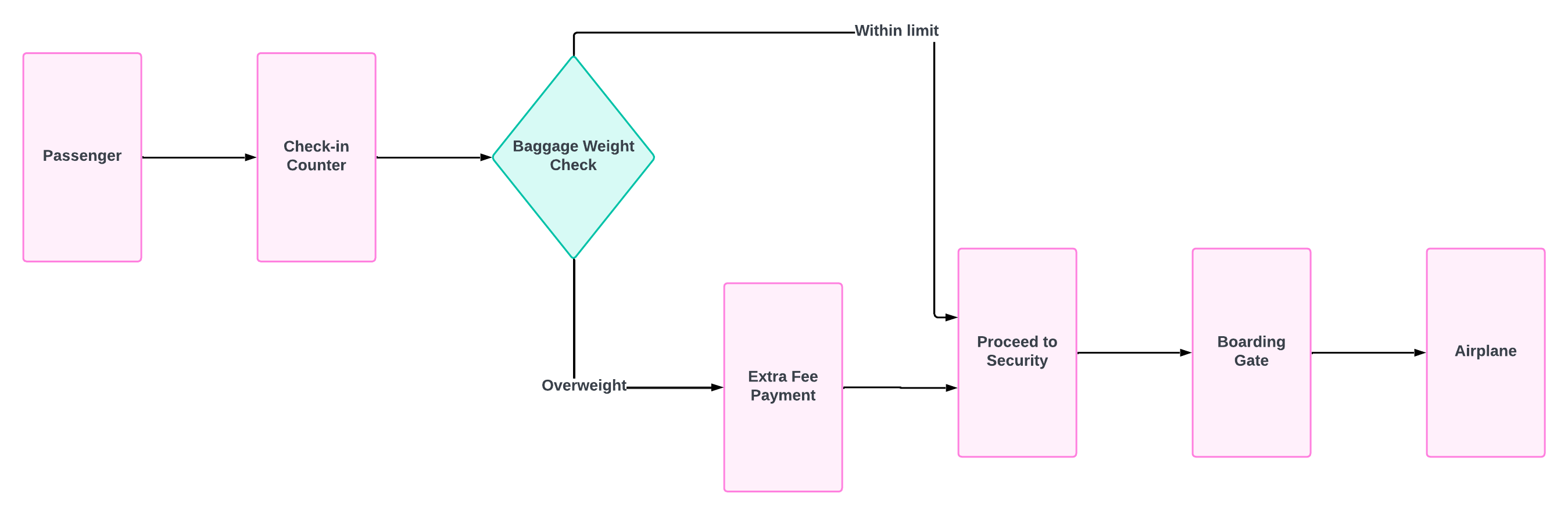}} \caption{ATRS Workflow}
\label{fig:agh}
\end{figure}

The Baggage Module in the Air Travel Recommender System (ATRS) is motivated by the need to enhance the travel experience for users. As illustrated in Table \ref{table:I} and Table \ref{table:II}, universally prohibited and country-specific prohibited item categories are identified. User interactions with baggage items play a crucial role in various contexts such as searching, buying, visiting, and more. The recommender system is conceptualized as an algorithm determining the probability that a traveler (user) would be interested in interacting with a particular item or service. Additionally, airline baggage appearance and transportability detection take place in the airport \cite{gao2021airline}. Quality assessment is also required in the baggage handling system \cite{rezaei2018quality}, and airport quality is utilized for the baggage handling system \cite{shojaei2018airports}. Based on these insights, the idea for personalized air travel baggage recommendations was developed. Originally introduced to tackle information overload challenges stemming from extensive catalogs, recommender systems aim to provide users with personalized recommendations \cite{dadoun2023recommender}, \cite{ricci2015user}. The Baggage Module specifically aims to narrow down the search for users, presenting a tailored subset of baggage items that align with the individual preferences and needs of the traveler. For example, Figure \ref{fig:ag} illustrates the challenge faced by travelers during packing, often juggling numerous items while pressed for time. This highlights the need for an Air Travel Baggage Recommender System (ATRS) to assist in selecting essential items efficiently. After packing, delays at the check-in counter due to baggage weight issues can be avoided using ATRS, ensuring that travelers proceed smoothly to security and boarding gates without incurring extra fees. Figure \ref{fig:agh} further illustrates the general challenges passengers encounter, emphasizing the role of ATRS in addressing these issues.
\subsection{Contributions}
Here, we outline the novel contributions of our research:
\begin{enumerate}
    \item \textbf{Identification of Constraints in ATRS Problem}: We identify and elucidate the specific constraints associated with the Air Travel Baggage Recommendation System (ATRS) problem.
    \item \textbf{Identification of Research Gaps}: We identified existing research gaps within the domain of ATRS, providing insights into areas requiring further investigation and development.
    \item \textbf{Analysis of Technological Improvements}: We analyze and document the current technological advancements relevant to the ATRS problem, assessing their impact and applicability. 
    \item \textbf{Resolution of Common Data Creation Challenges}: We propose solutions to common challenges encountered during the creation and preparation of datasets for ATRS applications.  
    \item \textbf{Utilization of Web Scraping Methods}: We employ web scraping techniques and tools to generate a comprehensive dataset tailored to the ATRS domain.
    \item \textbf{Introduction of a Novel Baggage Recommendation System}: We introduce a novel approach to baggage recommendation using FastText embeddings and Association Rule Mining (ARM), presenting a phased methodology for system development.
    \item \textbf{Phase-I: FastText Embeddings and Similarity Scores}: We implement Phase-I of our recommendation system, leveraging FastText embeddings to compute similarity scores between user inputs and existing items.
    \item \textbf{Phase-II: Content-Based Recommendations with User History}: Phase-II extends the system to incorporate user search history and timestamps into content-based recommendations.
    \item \textbf{Phase-III: ARM on User Search History for Dataset Creation}: We develop the ATRS vocabulary or dataset using Association Rule Mining (ARM) applied to user search history.
    \item \textbf{Phase-IV: FastText and ARM-Based Recommendations}: In Phase-IV, we utilize the ATRS vocabulary to generate recommendations using FastText embeddings and ARM.
    \item \textbf{Evaluation Metrics}: We evaluate our model using metrics such as support, confidence, lift, leverage, and convictions to assess its performance and effectiveness.
    \item \textbf{Comparison with Market-Basket Analysis Data}: We compare our ATRS model with market-basket analysis data to validate its efficacy and analyze its performance against established benchmarks.
\end{enumerate}
This paper organizes follow to introduces a Air travel baggage recommendation engine that advises travelers on specific items, recommending whether to pack them in check-in or cabin baggage, and identifying prohibited items. In addition, it recommends the top similar items related to the searched item and
whether to carry it in check-in or cabin baggage. The manuscript is structured as follows: Section II describes related work of the Air Travel Recommender Systems (ATRS) and the concept of ATRS. Section III describes ATRS Terminology and Taxonomy of terms. Section IV describes the methodology, Section V describes data description and visualization. section VI describes the proposed framework, Section VII describes the analysis of experimental results, and Section VIII describes Discussion and Concluding Remarks.
\section{Related work of the Air Travel Baggage Recommender Systems} 
\begin{longtable}{p{4cm}p{6cm}p{4cm}}
    \caption{Research in the Context of Air Travel Baggage Recommender Systems}
    \label{tab:air_travel_baggage_research} \\
    \toprule
    \textbf{Research Paper} &\textbf{Methodology} & \textbf{Study Scope} \\
    \midrule
    \endfirsthead
    \multicolumn{3}{c}%
    {{\tablename\ \thetable{} -- continued from previous page}} \\
    \toprule
    \textbf{Research Paper} & \textbf{Methodology}& \textbf{Study Scope} \\
    \midrule
    \endhead
    \midrule \multicolumn{3}{r}{{Continued on next page}} \\
    \endfoot
    \bottomrule
    \endlastfoot
    Smith et al. \cite{lever2018collaborative} & Machine Learning (Collaborative Filtering) & Global airline passengers; personalized packing recommendations based on historical data and traveler preferences \\
    Johnson and Lee (2020) \cite{qurashi2020document} & Natural Language Processing (NLP) & Domestic travelers; semantic analysis of packing lists and item recommendations \\
    Chen and Wang (2019) \cite{chen2019data}& Optimization Algorithms (Genetic Algorithms) & International business travelers; packing optimization considering weight restrictions and travel duration \\
    Liu and Zhang (2017) \cite{liu2017research} & Data Mining (Association Rule Mining) & Frequent flyer program members; mining associations between travel destinations and preferred baggage items \\
    Brown et al. (2021) \cite{carlini2021extracting} & Hybrid Approach (Machine Learning + Expert System) and LLM (Large Language Model) & Leisure travelers; training data from LLM and integrating expert advice with machine learning models for personalized packing recommendations \\
    Bernardis et al. (2022) \cite{bernardis2022nfc} & Hybrid-based Recommender Systems & In ATRS, recommending new items (cold-start) is challenging but crucial given their frequent introduction. ATRS faces difficulty in recommending these items due to the absence of prior preferences, requiring innovative approaches for effective recommendations \\
    Yang et al. (2023) \cite{song2023toward} & AI and IoT using Apriori algorithm & In ATRS designed to enhance the travel experience for air passengers by offering personalized packing recommendations, travel insights, and itinerary suggestions \\
    Alves et al. (2023) \cite{alves2023group} & Recommender Systems & Explores group recommender systems and addresses challenges like the cold start problem, relevant for collaborative baggage recommendations in ATRS \\
    Radek et al. (2024) \cite{pelanek2024personalized} & Recommender Systems & In Air Travel Recommender Systems (ATRS), personalization enhances travel by tailoring baggage recommendations to individual preferences. ATRS uses advanced algorithms to predict user preferences, offering personalized packing suggestions based on historical data \\
    Barile et al. (2023) \cite{barile2023evaluating} & Explainable AI & Social choice aggregation strategies provide explainable group recommendation methods, yet selecting the optimal strategy for specific groups is challenging, particularly in Air Travel Recommender Systems (ATRS), which must balance diverse traveler preferences for transparent and coherent outcomes \\
\end{longtable}
Recommender systems, particularly in the air travel industry such as Air Travel Recommender Systems (ATRS), play a vital role in predicting user preferences and offering personalized recommendations to enhance user experience \cite{schafer2001commerce}, \cite{esmaeili2020novel}. These systems address challenges such as popularity bias, which can lead to item starvation by recommending only popular items \cite{gupta2023eqbal}. ATRS aims to provide diverse and balanced recommendations tailored to individual traveler needs within the dynamic context of air travel.
Various methodologies are employed in recommender systems, including content-based filtering, collaborative filtering, knowledge-based filtering, and context-aware filtering \cite{esmaeili2011personalizing}, \cite{ravi2022comparative}, \cite{esmaeili2012hybrid}. These methods tackle challenges like the cold start problem and new item sets, often using hybrid approaches to enhance overall performance \cite{burke2002hybrid}, \cite{alves2023group}. Evaluating recommender systems involves metrics derived from precision-based measures, ensuring recommendations are diverse and not biased toward popular choices \cite{mendoza2020evaluating}.
In the tourism industry, recommender systems provide personalized suggestions for various goods and experiences, leveraging strategies such as agent-based collaboration and hybrid recommendation approaches \cite{ricci2022recommender}, \cite{borras2014intelligent}, \cite{lucas2013hybrid}. Session-aware recommender systems personalize recommendations based on user interactions across multiple sessions, optimizing recommendations for diverse and accurate suggestions \cite{khurana2023session}.
Integrating smart city concepts into air travel baggage systems offers benefits such as streamlined processes, enhanced tracking capabilities, optimized resource allocation, and improved security \cite{zanella2014internet}, \cite{mehmood2017internet}, \cite{qian2019internet}. Personalized itinerary recommendation optimizes air travel baggage handling by considering factors like flight schedules and baggage pickup locations, leveraging real-time data and analytics to enhance passenger experience \cite{halder2022efficient}, \cite{sarkar2023tourism}, \cite{zheng2018tourism}.
Recommender systems for travel baggage items use implicit and explicit feedback techniques, demographic-based filtering, and traveler similarity analysis to tailor recommendations based on individual preferences and characteristics \cite{burke2000knowledge}, \cite{pazzani1999framework}.
This comprehensive overview emphasizes the importance of recommender systems, especially Air Travel Recommender Systems (ATRS), in addressing personalized travel needs and enhancing user satisfaction within the air travel domain. Advanced methodologies and smart city integration further optimize baggage handling processes and passenger experiences, enhancing overall travel efficiency. The study mentioned above highlight diverse approaches and innovations in the design and implementation of air travel baggage recommender systems, collectively contributing to the advancement of personalized and efficient travel experiences for airline passengers.
\section{ATRS Terminology and Taxonomy of Terms}
Before starting to implement an ATRS, it's a good idea to consider the kind of system you want to roll out of the garage. To achieve this, a framework for investigating and defining recommender systems should be created. The taxonomy used to describe a system includes the following dimensions: domain, aim, context, personalization level, who's opinions, privacy and trustworthiness, interfaces, and algorithms. Examining each of the variables \cite{riedl2002word}. 
\subsection{How ATRS is useful}
ATRS can leverage the proposed baggage item recommendation model's advanced methodology, combining FastText word embeddings and Association Rule Mining, to enhance personalized suggestions for travelers, optimizing their baggage choices and overall travel experience. This contributes to the broader societal shift toward automation-driven advancements in the aviation sector.
Stages of Advice ``Do not carry".
Prohibited Items Universally Table \ref{table:I}. Items prohibited by destination or country-wide Table \ref{table:II}.
Items are not suitable based on airplane compatibility (weight or size restrictions).
\begin{table}[ht]
\caption{Universally prohibited items}
\label{table:I}
\begin{tabular}{|c|c|}
\hline
\textbf{Items} & \makecell{\textbf{Prohibited items list universally}}\\
 \hline
 \textbf{Explosives}&  \makecell{ These include blasting caps, fireworks, dynamite, and fireworks.}    \\
 \hline
\textbf{Flammable} & \makecell{Gasoline, lighter fluid and aerosol sprays (with the exception \\of certain personal care products).}    \\
 \hline
\textbf{Compressed gases} & \makecell{These include butane, propane, and oxygen cylinders\\ (apart from portable oxygen concentrators \\used for medical reasons).}\\
 \hline
 \textbf{Firearms and weapons} & \makecell{These include actual firearms,\\ ammunition, explosives, and realistic\\looking weapon reproductions.}\\
 \hline
\textbf{Sharp objects}& \makecell{These include razor blades, box cutters,\\ and knives but not butter knives\\ or knives with circular\\ blades or plastic handles.}\\
 \hline
 \textbf{Self defense}  & \makecell{These includes \\ Stun guns, mace, and pepper spray.}\\
 \hline
 \textbf{Hazardous materials} & \makecell{ These include corrosives, toxins,\\ radioactive substances, and infectious agents.}\\
 \hline
 \textbf{Liquids and gels} & \makecell{Unless contained in a transparent, resealable \\plastic bag, in containers that exceed \\the maximum permitted limit \\(about 3.4 ounces or 100 millilitres per container).}\\
 \hline
 \textbf{Batteries} & \makecell{It may be forbidden or restricted to use \\spare lithium batteries\\ that have a capacity that exceeds certain thresholds \\(often more than 100 watt-hours).
}\\
 \hline
\end{tabular}
\end{table}
\begin{table}[ht]
\caption{Country wide prohibited items.}
\label{table:II}
\begin{tabular}{|c|c|}
\hline
\textbf{Items} & \makecell{\textbf{Items prohibited}}\\
 \hline
 \textbf{Weapons and firearms}&  \makecell{  In general, guns, explosives,\\ and other weapons are not \\allowed in public areas, \\including on aeroplanes.}    \\
 \hline
\textbf{Narcotics:} & \makecell{ It is totally forbidden to use\\ any illegal drugs or narcotics.}    \\
 \hline
\textbf{Counterfeit goods} & \makecell{Items that violate intellectual\\ property rights,\\ such as knockoffs of popular brands \\or designer goods, are typically forbidden.}\\
 \hline
\textbf{\makecell{Obscene or offensive materials}}& \makecell{Obscene, pornographic,\\ or offensive materials may not be allowed.}\\
 \hline
 \textbf{\makecell{Wildlife products \\and endangered species}}  & \makecell{It is forbidden to use ivory,\\ some animal hides, or\\ unusual animal goods that are created \\from, or derived from,\\ endangered species.}\\
 \hline
 \textbf{\makecell{Archaeological and cultural artifacts}} & \makecell{ Several nations prohibit\\ the import or export of \\antiquities, cultural artifacts,\\ or both without the required\\ documentation.}\\
 \hline
 \textbf{Agriculture} & \makecell{To stop the spread of \\illnesses or pests, some nations have stringent\\ laws governing the importation of\\ agricultural products like \\fruits, vegetables, plants, and animal products.}\\
 \hline
\end{tabular}
\end{table}
And the following passage describes about the problem statement.
\subsection{Problem statement}
In the following we are defining the problem statement of a multi-tiered solution for personalized baggage item recommendation system.\\
Let $I = \{i_1, i_2, \ldots, i_n\}$ be the set of all baggage items, where $n$ is the total number of items available.\\
Let $U = \{u_1, u_2, \ldots, u_m\}$ be the set of all users, where $m$ is the total number of users.\\
Let $T = \{t_1, t_2, \ldots, t_p\}$ be the set of all trips or travel instances, where $p$ is the total number of trips.\\
Each baggage item $i$ is characterized by a set of attributes $A_i = \{a_{i1}, a_{i2}, \ldots, a_{ik}\}$, where $k$ is the number of attributes describing each item. Each user $u$ has associated preferences $P_u = \{p_{u1}, p_{u2}, \ldots, p_{ur}\}$, where $r$ is the number of preference factors considered. Each trip $t$ is defined by parameters such as destination, duration, and purpose.
\subsection{Objectives and Constraints of ATRS} In this passage
we are discussing regarding the ATRS objectives and constraints.
\textbf{Objectives in ATRS}: The primary objective of the Air Travel Recommender System (ATRS) is to recommend a set of baggage items $R_{u,t}$ that optimally align with a user $u$'s preferences and trip requirements $t$.\\
\textbf{Constraints in ATRS}: In this passage, we explore the specific constraints that influence the design and operation of the Air Travel Recommender System (ATRS).
\begin{enumerate}
    \item \textbf{Relevance Constraint}: Recommended items should be relevant to the user's preferences and trip context.
    \item \textbf{Size and Weight Constraint}: Recommended items should adhere to size and weight restrictions imposed by airlines or travel regulations.
    \item \textbf{Personalization Constraint}: Recommendations should be personalized based on the user's past behavior and preferences.
    \item \textbf{Association Rule Constraint}: Recommendations should leverage association rules mined from historical user-item interactions.
    \item \textbf{Efficiency Constraint}: The recommendation process should be computationally efficient to handle large datasets and real-time user requests.
\end{enumerate}
\textbf{Mathematical Representation}:
The recommendation process can be formulated as an optimization problem:
\[
\max_{R_{u,t}} f(R_{u,t})
\]
subject to:
\begin{align*}
    &\text{Relevance Constraint}: \quad R_{u,t} \subseteq I \\
    &\text{Size and Weight Constraint}: \quad \sum_{i \in R_{u,t}} (size_i \times quantity_i) \leq \text{MaxSize} \\
    &\text{Personalization Constraint}: \quad R_{u,t} \subseteq P_u \\
    &\text{Association Rule Constraint}: \quad R_{u,t} \text{ satisfies mined association rules} \\
    &\text{Efficiency Constraint}: \quad \text{Time complexity} \leq \text{Threshold}
\end{align*}
Where:
\begin{enumerate}
    \item $f(R_{u,t})$ represents the utility or satisfaction derived from the recommended items $R_{u,t}$.
    \item $size_i$ and $quantity_i$ denote the size and quantity of each item $i$ in the recommendation set.
    \item \text{MaxSize} represents the maximum allowable size for baggage.
    \item \text{Time complexity} refers to the computational complexity of the recommendation process.
    \item \text{Threshold} is a predefined threshold for acceptable computation time.
\end{enumerate}
This mathematical formulation captures the essence of the problem, outlining the objective, constraints, and decision variables involved in recommending baggage items for travelers.
\subsection*{Baggage Constraints Overview}
Table \ref{tab:baggage-constraints}  provides an overview of the baggage constraints for popular Indian airlines.
\begin{table}[htbp]
    \centering
    \caption{Baggage Constraints for Indian Airlines}
    \label{tab:baggage-constraints}
    \begin{tabular}{|p{2cm}|p{3cm}|p{4cm}|p{4cm}|}
        \toprule
        \textbf{Airline} & \textbf{Cabin/Carry on Baggage (kg)}& \textbf{Cabin/Carry on Baggage (dimensions)} & \textbf{Check-in Baggage (free allowance)} \\
        \midrule
        IndiGo            & Up to 7                      & 55 cm x 35 cm x 25 cm                & 15 kg to 30 kg                             \\
        Air India         & Up to 8                      & 55 cm x 35 cm x 25 cm                & Varies (25 kg to 35 kg)                     \\
        SpiceJet          & Up to 7                      & 55 cm x 40 cm x 20 cm                & 15 kg to 30 kg                             \\
        Vistara           & Up to 7 (Economy), 12 (Premium Economy) & 55 cm x 40 cm x 20 cm     & 15 kg to 30 kg                             \\
        \bottomrule
    \end{tabular}
\end{table}
Table \ref{tab:tsa-baggage-constraints} provides the TSA baggage constraints for carry-on and checked baggage:
\subsection*{TSA Baggage Constraints Overview}
Table \ref{tab:tsa-baggage-constraints} encapsulates the key baggage constraints enforced by the Transportation Security Administration (TSA) for air travel within the United States.
\begin{table}[htbp]
    \centering
    \caption{TSA Baggage Constraints for Air Travel}
    \label{tab:tsa-baggage-constraints}
    \begin{tabular}{|p{4cm}|p{9cm}|}
        \hline
        \textbf{Baggage Type} & \textbf{Constraints} \\
        \hline
        \textbf{Carry-On Baggage} & 
        \begin{itemize}
            \item \textbf{Size Restrictions:} Maximum dimensions of 22 inches x 14 inches x 9 inches (including wheels and handles).
            \item \textbf{Weight Limit:} No specific weight limit, but must be light enough for passenger to stow overhead unassisted.
            \item \textbf{Prohibited Items:} Certain items like sharp objects, firearms, explosives, and liquids over 3.4 ounces are not allowed.
        \end{itemize} \\
        \hline
        \textbf{Checked Baggage} & 
        \begin{itemize}
            \item \textbf{Size and Weight Limits:} Maximum weight of 50 pounds (23 kilograms) and linear dimensions (length + width + height) not exceeding 62 inches.
            \item \textbf{Prohibited Items:} Similar restrictions as carry-on baggage, with additional limitations on certain chemicals and substances.
        \end{itemize} \\
        \hline
    \end{tabular}
\end{table}
Travelers should always consult the TSA website or their airline for the most up-to-date information and specific guidelines before packing for a trip.
\subsection{Research Gaps in Air Travel Baggage Recommendation Systems}
Baggage recommendation systems play a crucial role in enhancing the travel experience by providing personalized suggestions for packing items. However, several research gaps exist in the current state of air travel baggage recommendation systems:
\subsubsection{Contextual Understanding}
Existing systems often lack robust contextual understanding, failing to consider specific travel scenarios, destination climates, or individual preferences. Future research should focus on enhancing context-awareness to deliver more tailored recommendations \cite{ricci2010introduction}. Mathematically, we can represent the contextual understanding in Air Travel Recommender Systems (ATRS) as follows:
Let \( S \) denote the set of all travel scenarios, \( C \) represent the set of destination climates, and \( P \) denote the set of individual preferences. An ATRS system aims to enhance context-awareness by incorporating these contextual factors into its recommendation function \( R(u) \), which provides personalized recommendations for a user \( u \) based on their context:
\[ R(u) = f(S_u, C_u, P_u) \]
where:
\begin{itemize}
    \item \( S_u \in S \) represents the specific travel scenario of user \( u \),
    \item \( C_u \in C \) represents the destination climate preference of user \( u \),
    \item \( P_u \in P \) represents the individual preferences of user \( u \), and
    \item \( f(\cdot) \) is a recommendation function that considers the user's context to generate tailored recommendations.
\end{itemize}
Future research in ATRS should aim to optimize \( f(\cdot) \) by leveraging advanced algorithms and techniques to enhance context-awareness and deliver more tailored recommendations that align with specific travel scenarios, destination climates, and individual preferences. Existing systems often lack robust contextual understanding, failing to consider specific travel scenarios, destination climates, or individual preferences. Future research should focus on enhancing context-awareness to deliver more tailored recommendations\cite{ricci2010introduction}.
\subsubsection{Integration of Real-Time Data}
Current systems may not effectively incorporate real-time data updates, such as flight delays, weather conditions, and security regulations. Research is needed to explore methods for integrating dynamic data sources to improve recommendation accuracy \cite{moens2014mining}, \cite{burke2002hybrid}.
Mathematically, the integration of real-time data into an ATRS system can be represented as follows:
Let \( D(t) \) represent the set of dynamic data sources available at time \( t \), including flight delay information \( \text{FD}(t) \), weather conditions \( \text{WC}(t) \), and security regulations \( \text{SR}(t) \).
An ATRS system aims to incorporate these real-time data sources into its recommendation function \( R(u) \), which provides personalized recommendations for a user \( u \) based on the current context \( t \):
\[ R(u, t) = f(u, D(t)) \]
where:
\begin{itemize}
    \item \( u \) represents the user for whom recommendations are generated,
    \item \( D(t) = \{ \text{FD}(t), \text{WC}(t), \text{SR}(t) \} \) denotes the real-time data available at time \( t \),
    \item \( f(\cdot) \) is a recommendation function that integrates user information and real-time data to generate personalized recommendations.
\end{itemize}
Future research in ATRS should focus on developing advanced algorithms and techniques within \( f(\cdot) \) to effectively utilize real-time data sources \( D(t) \) and improve recommendation accuracy based on up-to-date travel conditions.
\subsubsection{Personalization and User Profiling}
Mathematically, personalization and user profiling in ATRS can be represented as follows:
Let \( U \) denote the set of all users of the ATRS system, where each user \( u \in U \) is characterized by various attributes such as demographics (\( \text{Dem}(u) \)), travel history (\( \text{TH}(u) \)), and behavioral patterns (\( \text{BP}(u) \)).
The goal of ATRS is to personalize recommendations for each user \( u \) based on their individual profile. This is achieved through a personalized recommendation function \( R(u) \) that incorporates user-specific attributes:
\[ R(u) = f(\text{Dem}(u), \text{TH}(u), \text{BP}(u)) \]
where:
\begin{itemize}
    \item \( \text{Dem}(u) \) represents the demographic information of user \( u \),
    \item \( \text{TH}(u) \) represents the travel history of user \( u \),
    \item \( \text{BP}(u) \) represents the behavioral patterns of user \( u \),
    \item \( f(\cdot) \) is a personalized recommendation function that leverages user-specific attributes to generate tailored recommendations.
\end{itemize}
Future research in ATRS should focus on developing sophisticated algorithms within \( f(\cdot) \) to accurately capture and model user preferences based on diverse user attributes, thereby enhancing the personalization and effectiveness of the recommendation system.
While attempts are made to personalize recommendations, there is room for improvement in user profiling and preference modeling. Research gaps exist in developing advanced techniques to capture nuanced traveler preferences based on demographics, travel history, and behavioral patterns \cite{bobadilla2013recommender}, \cite{zukerman2001predictive}, \cite{ricci2015user}.
\subsubsection{Multi-Modal Recommendations}
Baggage recommendations predominantly focus on packing items but could be extended to other aspects of the travel experience, such as ground transportation, accommodations, or local attractions. Future research could explore multi-modal recommendation strategies to enhance overall travel satisfaction \cite{ricci2010introduction}, \cite{zukerman2001predictive}.
Mathematically, multi-modal recommendations in ATRS can be represented as follows:
Let \( I \) denote the set of all relevant travel items, \( T \) denote the set of travel activities, and \( P \) denote the set of travel preferences.
The goal of ATRS is to provide recommendations across multiple travel modalities, integrating packing items (\( I \)), travel activities (\( T \)), and traveler preferences (\( P \)). This is achieved through a multi-modal recommendation function \( R(I, T, P) \) that encompasses diverse aspects of the travel experience:
\[ R(I, T, P) = f(I, T, P) \]
where:
\begin{itemize}
    \item \( I \) represents packing items,
    \item \( T \) represents travel activities,
    \item \( P \) represents traveler preferences,
    \item \( f(\cdot) \) is a multi-modal recommendation function that combines information from packing items, travel activities, and traveler preferences to enhance overall travel satisfaction.
\end{itemize}
Future research in ATRS should focus on developing sophisticated algorithms within \( f(\cdot) \) to seamlessly integrate multi-modal recommendations, providing travelers with comprehensive guidance and enhancing their overall travel experience.
\subsubsection{User Interaction and Feedback}
Current systems often under utilize user interactions and feedback to refine recommendation quality. Research should investigate interactive interfaces that facilitate seamless feedback loops, enabling users to personalize recommendations based on evolving needs \cite{zhao2013interactive}.
Mathematically, user interaction and feedback in ATRS can be represented as follows:
Let \( U \) denote the set of users, \( I \) denote the set of all relevant travel items, and \( R \) denote the set of recommendations.
The goal of ATRS is to leverage user interactions and feedback to improve recommendation quality through a feedback loop mechanism \( F(U, I, R) \):
\[ F(U, I, R) = g(U, I, R) \]
where:
\begin{itemize}
    \item \( U \) represents users providing feedback,
    \item \( I \) represents relevant travel items,
    \item \( R \) represents recommendations generated by ATRS,
    \item \( g(\cdot) \) is a function that incorporates user feedback to refine recommendations and enhance user satisfaction.
\end{itemize}
Future research in ATRS should focus on developing interactive interfaces and feedback mechanisms within \( F(\cdot) \) to optimize user interaction and improve recommendation quality based on evolving user needs.
\subsubsection{Evaluation Metrics and Benchmarks}
Establishing standardized evaluation metrics and benchmarks for baggage recommendation systems remains a challenge. Future research should develop objective measures to assess the effectiveness, relevance, and user satisfaction of recommendations across diverse travel scenarios \cite{cremonesi2010performance}.
Mathematically, evaluation metrics and benchmarks in ATRS can be represented as follows:
Let \( R \) denote the set of recommendations generated by ATRS, \( U \) denote the set of users, and \( S \) denote the set of scenarios representing diverse travel contexts. The goal of ATRS is to define evaluation functions \( E(R, U, S) \) that quantify the quality of recommendations based on user satisfaction:
\[ E(R, U, S) = f(R, U, S) \]
where:
\begin{itemize}
    \item \( R \) represents recommendations generated by ATRS,
    \item \( U \) represents users evaluating recommendations,
    \item \( S \) represents diverse travel scenarios,
    \item \( f(\cdot) \) is a function that computes evaluation metrics based on user feedback and scenario relevance.
\end{itemize}
Future research in ATRS should focus on developing standardized evaluation metrics and benchmarks within \( E(\cdot) \) to objectively assess recommendation effectiveness and relevance across diverse travel scenarios.
\subsubsection{Privacy and Trust}
Ensuring user privacy and fostering trust in baggage recommendation systems are critical concerns. Research gaps exist in developing privacy-preserving techniques and transparent recommendation approaches that prioritize user control over data sharing \cite{wang2018toward}. Mathematically, privacy and trust considerations in ATRS can be represented as follows:
Let \( D \) denote the dataset used by ATRS and \( U \) denote the set of users interacting with the system. The goal of ATRS is to define privacy-preserving functions \( P(D, U) \) that ensure user data protection and trust:
\[ P(D, U) = f(D, U) \]
where:
\begin{itemize}
    \item \( D \) represents the dataset used by ATRS,
    \item \( U \) represents users interacting with the system,
    \item \( f(\cdot) \) is a function that implements privacy-preserving techniques to protect user data and ensure trust.
\end{itemize}
Future research in ATRS should focus on developing privacy-preserving techniques within \( P(\cdot) \) to enhance user trust and confidence in baggage recommendation systems.
\subsubsection{Adaptation to Cultural and Regional Factors}
Baggage preferences vary across cultures and regions, which are often not adequately accounted for in existing systems. Future research should explore methods to adapt recommendations to cultural norms, local regulations, and regional travel customs \cite{jung2018cross}.
Mathematically, adaptation to cultural and regional factors in ATRS can be represented as follows:
Let \( B \) denote the set of baggage items, \( C \) denote cultural norms, and \( R \) denote regional travel customs. The goal of ATRS is to define adaptation functions \( A(B, C, R) \) that tailor recommendations based on cultural and regional factors:
\[ A(B, C, R) = f(B, C, R) \]
where:
\begin{itemize}
    \item \( B \) represents the set of baggage items,
    \item \( C \) represents cultural norms,
    \item \( R \) represents regional travel customs,
    \item \( f(\cdot) \) is a function that adapts recommendations based on cultural and regional factors.
\end{itemize}
Future research in ATRS should focus on developing \( f(\cdot) \) to effectively adapt recommendations to diverse cultural and regional preferences. Addressing these research gaps will contribute to the advancement of air travel baggage recommendation systems, leading to more sophisticated and user-centric solutions that enhance convenience, efficiency, and satisfaction for travelers. Ongoing research efforts in these areas are essential for improving the ATRS in this domain.
\subsection{Advanced technological improvement in Air travel baggage recommendation system}
The future advancements in air travel baggage recommendation systems represent substantial improvements over the existing ATRS, introducing innovative features and capabilities that significantly enhance user experience and system performance.
\subsubsection{Enhanced Personalization}
One of the key real improvements is the enhanced level of personalization in baggage recommendations. Unlike traditional systems that provide generic suggestions, ATRS systems leverage advanced machine learning techniques to analyze traveler preferences, historical data, and contextual factors (e.g., destination, travel purpose) to deliver tailored recommendations. This personalized approach ensures that recommendations closely align with individual traveler needs and preferences \cite{lim2018personalized}, \cite{lim2016personalized}, \cite{zukerman2001predictive}.
\subsubsection{Real-Time Adaptability}
Modern baggage recommendation systems excel in real-time adaptability, allowing them to dynamically adjust recommendations based on changing travel circumstances. By integrating with live flight data and itinerary updates, these systems can provide up-to-the-minute suggestions, accommodating flight delays, weather changes, or itinerary modifications. This real-time adaptability significantly enhances user convenience and flexibility during travel \cite{bin2019travel}, \cite{ricci2015user}.
\subsubsection{Integration of Advanced Technologies}
ATRS systems integrate cutting-edge technologies such as deep learning models and natural language processing (NLP), 5G, XAI (Explainable AI) to analyze vast amounts of traveler data efficiently. By harnessing these advanced techniques, systems can generate more accurate and contextually relevant recommendations, leading to higher user satisfaction and engagement \cite{wang2020realizing}, \cite{cepeda2022deep}.
\subsubsection{Improved User Interface and Experience}
Another notable improvement is the focus on enhancing the user interface (UI) and overall user experience (UX). Modern systems prioritize intuitive design, visually appealing interfaces, and seamless integration with travel apps and platforms. This improved UI/UX translates into higher user adoption rates and improved usability across diverse user demographics \cite{afzal2018personalization}, \cite{ricci2015user}.
\subsubsection{Context-Aware Recommendations}
ATRS systems leverage contextual awareness to deliver more relevant recommendations. By considering factors such as travel purpose, weather conditions, and cultural norms, these systems can suggest appropriate items for different types of trips (e.g., business travel, leisure travel, family vacations). This context-aware approach contributes to more informed and practical recommendations \cite{renjith2020extensive}.
\subsubsection{Data Security and Privacy Measures}
Advanced baggage recommendation systems prioritize robust data security and privacy measures to protect user information. By implementing encryption protocols, secure data storage, and compliance with privacy regulations (e.g., GDPR), these systems ensure the confidentiality and integrity of traveler data, fostering trust and reliability among  \cite{jiang2019towards}.
\subsubsection{Scalability and Performance Optimization}
Modern systems are designed for scalability and optimized performance, capable of handling large volumes of data and user requests. By leveraging cloud computing and distributed architectures, these systems can deliver responsive recommendations even during peak travel periods, ensuring reliable performance and scalability. In the real improvements in air travel baggage recommendation systems reflect a trans-formative evolution in personalized travel experiences. These advancements elevate the quality of service, optimize system performance, and contribute to the overall efficiency and satisfaction of travelers in the modern air travel \cite{sarwar2000analysis}.
\subsubsection{Future Technological Improvement} In the future ATRS we can use Machine learning, Deep Learning, Natural language Processing (NLP) and Explainability methods we can use in the ATRS domain for resolving the packing and scanning the baggage problems \cite{VANDERVELDEN2022102470}. The following section describes about methodology.
\section{Methodology}
In our data collection and analysis pipeline, we utilize web scraping techniques with tools such as Selenium WebDriver and BeautifulSoup to extract information from targeted websites. The acquired raw dataset, initially in CSV format, undergoes subsequent data preprocessing involving cleaning, transforming, and organizing to enhance suitability for modeling purposes \cite{kumar2014feature}, \cite{visalakshi2014literature}, \cite{al2020approaches}.
After preprocessing, feature selection is applied to the refined dataset for analysis, and the resulting insights are stored in a new CSV file. This final dataset, derived from model outputs, is then renamed as the `baggage dataset.' This systematic workflow ensures a comprehensive approach to data extraction, preparation, and utilization in our analytical endeavors. We are describing the step by step methodology part in the proposed frame work.
\subsection{Software Requirements}
The methodology relies on FastText, Python, and associated libraries for word embeddings, data processing, and programming, while incorporating Association Rule Mining algorithms, forming a robust software foundation for the baggage recommendation system.
\subsection{Importing the Libraries}
In our approach we are using the list of libraries like, pandas, os, numpy, keyedvectors, gensim models, appears to involve processing and analyzing textual data (potentially using word embeddings and cosine similarity), alongside general data manipulation and time-related operations. This combination of libraries suggests a focus on natural language processing (NLP) tasks, possibly for tasks like text similarity, recommendation systems, or other applications involving text data and embeddings. The data is loading into  data frame and tokenized the data. We Load the fast text pre-trained embeddings. And we apply cosine similarity based on similarity score we create unique categorised the dataset. And we build the Phase-I, Phase-II, Phase-III, Phase-IV recommendations for ATRS. In the proposed system we will explain in details. The following section describes about data description and visualization.
\section{Data Description and Visualization}
Creating benchmark data poses challenges when it's unavailable. In our research, we address common challenges by leveraging innovative techniques to simulate benchmark data, ensuring robust evaluation and analysis of our recommendation system.
\subsection{Common Key Challenges}
Here we are describing about common key challenges occurs at the time of data creation.
Let \( D \) represent the benchmark dataset for personalized baggage recommendation.\\
\textbf{Challenge:}
Creating a benchmark dataset (\( D \)) for personalized baggage recommendation when none is available is a major challenge.
\[ D = \emptyset \]
\textbf{Solution:}
To address this challenge, we employ web scraping techniques to gather data from air travel websites.
Let \( S \) denote the set of websites targeted for scraping.
\[ S = \{ \text{website}_1, \text{website}_2, \ldots, \text{website}_n \} \]
Let \( P \subseteq S \) be the subset of websites that permit scraping, and \( B = S \setminus P \) be the set of blocked websites.
\[ P = \{ \text{website}_i \in S \mid \text{permit scraping} \} \]
\[ B = \{ \text{website}_i \in S \mid \text{block scraping} \} \]
For permitted websites (\( \text{website}_i \in P \)), we utilize scraping tools to extract relevant data, denoted as \( D_P \).
\[ D_P = \bigcup_{\text{website}_i \in P} \text{scrape}(\text{website}_i) \]
For blocked websites (\( \text{website}_i \in B \)), we employ browser extensions to enable scraping, resulting in \( D_B \).
\[ D_B = \bigcup_{\text{website}_i \in B} \text{scrape\_with\_extension}(\text{website}_i) \]
Finally, we combine the data obtained from permitted and unblocked websites to form the benchmark dataset \( D \) for personalized baggage recommendation.
\[ D = D_P \cup D_B \]\\
\textbf{Challenge II:}
Extracting and compiling data from various airline websites while adhering to their specific guidelines and regulations presents a significant challenge in dataset creation for the Air Travel Recommender System (ATRS).
Let \( D \) denote the dataset for the Air Travel Recommender System.
\[ D = \emptyset \]\\
\textbf{Solution:}
To address this challenge, we thoroughly explore the guidelines provided by targeted airlines and regulatory bodies such as the Directorate General of Civil Aviation (DGCA). Each airline's guidelines typically include information on item names, carry-on allowances, checked baggage regulations, prohibited items, and descriptions.
Let \( A \) represent the set of targeted airline websites.
\[ A = \{ \text{airline}_1, \text{airline}_2, \ldots, \text{airline}_n \} \]
For each airline \( \text{airline}_i \in A \), we gather information according to their guidelines to create the dataset \( D \).
\[ D = \bigcup_{\text{airline}_i \in A} \text{extract\_data}(\text{airline}_i) \]
We incorporate this gathered information into the dataset \( D \), ensuring compliance with regulations and guidelines.
For items that cannot be scraped due to website restrictions, we manually include relevant data in the dataset \( D \).
The final dataset \( D \) is compiled with information extracted from various airline websites, providing valuable data for the Air Travel Recommender System.
\[ D = \text{final\_dataset} \]
\textbf{Challenge III:}
Handling data preprocessing issues related to loading data dependencies and pre-trained vectors poses significant challenges in the Air Travel Recommender System (ATRS) dataset creation.
Let \( D \) denote the dataset for the Air Travel Recommender System.
\[ D = \emptyset \]
\textbf{Solution:}
To address these challenges, we implement techniques such as tokenization and FastText embeddings for item names, handling variations like single-word, two-word, and multi-word item names.
Let \( I \) represent the set of all item names extracted from airline websites.
\[ I = \{ \text{item}_1, \text{item}_2, \ldots, \text{item}_n \} \]
We preprocess each item \( \text{item}_i \in I \) using tokenization and FastText embeddings to obtain vector representations \( \text{embedding}(\text{item}_i) \).
\[ \text{embedding}(\text{item}_i) = \text{FastText\_vector}(\text{tokenize}(\text{item}_i)) \]
This process resolves data dependency issues and loading errors associated with pre-trained embeddings.
We categorize items into top similar item categories based on these embeddings to streamline recommendation processes.
Let \( C \) be the set of item categories generated from embeddings.
\[ C = \{ \text{category}_1, \text{category}_2, \ldots, \text{category}_m \} \]
Each item \( \text{item}_i \) is assigned to one or more categories \( \text{category}_j \) based on similarity scores.
\[ \text{category}(\text{item}_i) = \arg\max_{\text{category}_j \in C} \text{similarity\_score}(\text{item}_i, \text{category}_j) \]
The final dataset \( D \) includes preprocessed item representations and their assigned categories, ready for use in the Air Travel Recommender System.
\[ D = \{ (\text{embedding}(\text{item}_i), \text{category}(\text{item}_i)) \mid \text{item}_i \in I \} \]\\
\textbf{Challenge IV:}
Establishing user preferences or search history for personalized recommendations poses a significant challenge, especially when no explicit user preferences exist in airline or Directorate General of Civil Aviation (DGCA) guidelines.
Let \( U \) represent the set of users interacting with the Air Travel Recommender System (ATRS).
\[ U = \{ \text{user}_1, \text{user}_2, \ldots, \text{user}_n \} \]\\
\textbf{Solution:}
To address this challenge, we employ content-based recommendations using cosine similarity.
Let \( I \) denote the set of all items in the baggage dataset.
\[ I = \{ \text{item}_1, \text{item}_2, \ldots, \text{item}_m \} \]
When a user \( \text{user}_i \) interacts with items in the dataset, we calculate the cosine similarity \( \text{similarity}(\text{user}_i, \text{item}_j) \) between the user profile and each item.
\[ \text{similarity}(\text{user}_i, \text{item}_j) = \frac{\text{user\_profile}(\text{user}_i) \cdot \text{item\_vector}(\text{item}_j)}{\| \text{user\_profile}(\text{user}_i) \| \| \text{item\_vector}(\text{item}_j) \|} \]
If a user interacts with an item \( \text{item}_j \) already existing in the dataset, we do not store these interactions in the user search history. However, if a user interacts with a new item \( \text{item}_k \) not present in the dataset, we record this interaction in the user search history \( H \) along with a timestamp \( t \).
\[ H = \{ (\text{user}_i, \text{item}_k, t) \} \]
This method enables us to collect and refine user preferences over time for personalized recommendations. To solve these challenges we proposed the framework for Multi-tiered approach for baggage recommendations incorporates these techniques to handle user preferences and provide personalized recommendations in the Air Travel Recommender System (ATRS).
\subsection{Data description}
To proposing the frame work or building the model the dataset is main requirements. Here the dataset is created using web scraping method. We web scrape the targeted airlines ``International Airlines and Airlines under Indian DGCA". First we explore the International airlines like TSA \footnote{\url{https://www.tsa.gov/travel/security-screening/whatcanibring/all}}and IATA and Indian Airlines like Airindia \footnote{\url{https://www.airindia.com/in/en/travel-information/baggage-guidelines/restricted-baggage.html}}, Indigo \footnote{\url{https://www.goindigo.in/baggage/dangerous-goods-policy.html}}, Vistara \footnote{\url{https://www.airvistara.com/in/en/travel-information/vistara-experience/on-ground/baggage/dangerous-goods}}, AirAsia \footnote{\url{https://support.airasia.com/s/article/What-items-are-prohibited-en?language=en_GB}}, Goindia \footnote{\url{https://www.goindigo.in/baggage/dangerous-goods-policy.html}} and Spice jet \footnote{\url{https://corporate.spicejet.com/airtravelbaggagefaq.aspx}}. And we observe the what are the items, are common in all, and what are the prohibited items, and what are the carry on, and what are the check in items. We create the dataset with relevant features like ``Item name," ``Carry on," ``Check-in," ``Prohibited," and ``Category" and ``ItemDescription". Initially we are scraping the airlines data we build the ATRS vocabulary contain 712 items. As follows the dataset visualization. 
\subsection{Dataset Visualization}
Here we are visualize the dataset.
\begin{figure}[H]
\centering
\fbox{\includegraphics[width=0.9\linewidth]{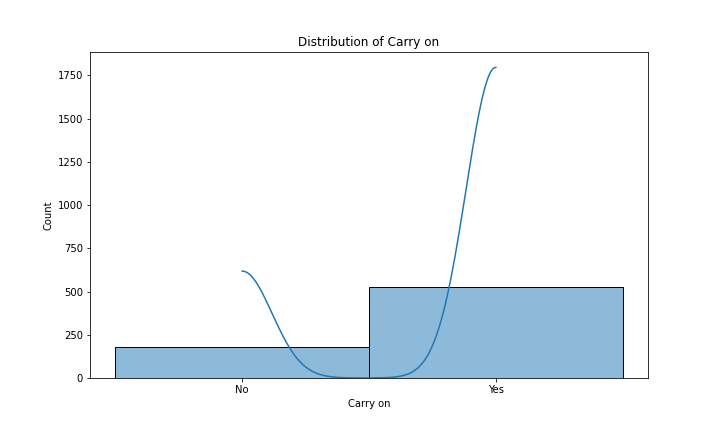}} \caption{Distribution of Carry on items}
\label{fig:A}
\end{figure} 
\begin{figure}[H]
\centering
\fbox{\includegraphics[width=0.9\linewidth]{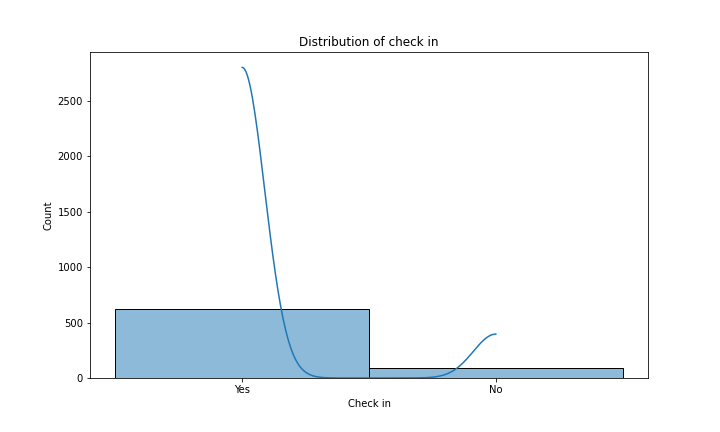}} \caption{Distribution of Check in items}
\label{fig:B}
\end{figure} 
\begin{figure}[H]
\centering
\fbox{\includegraphics[width=0.9\linewidth]{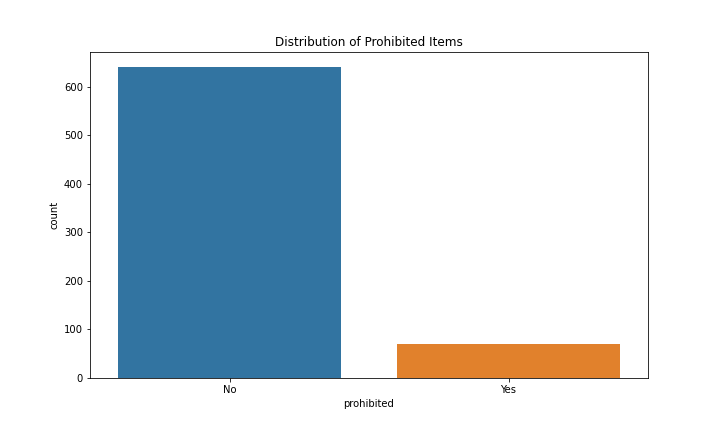}} \caption{Distribution of Prohibited items}
\label{fig:C}
\end{figure}  
\begin{figure}[H]
\centering
\fbox{\includegraphics[width=0.9\linewidth]{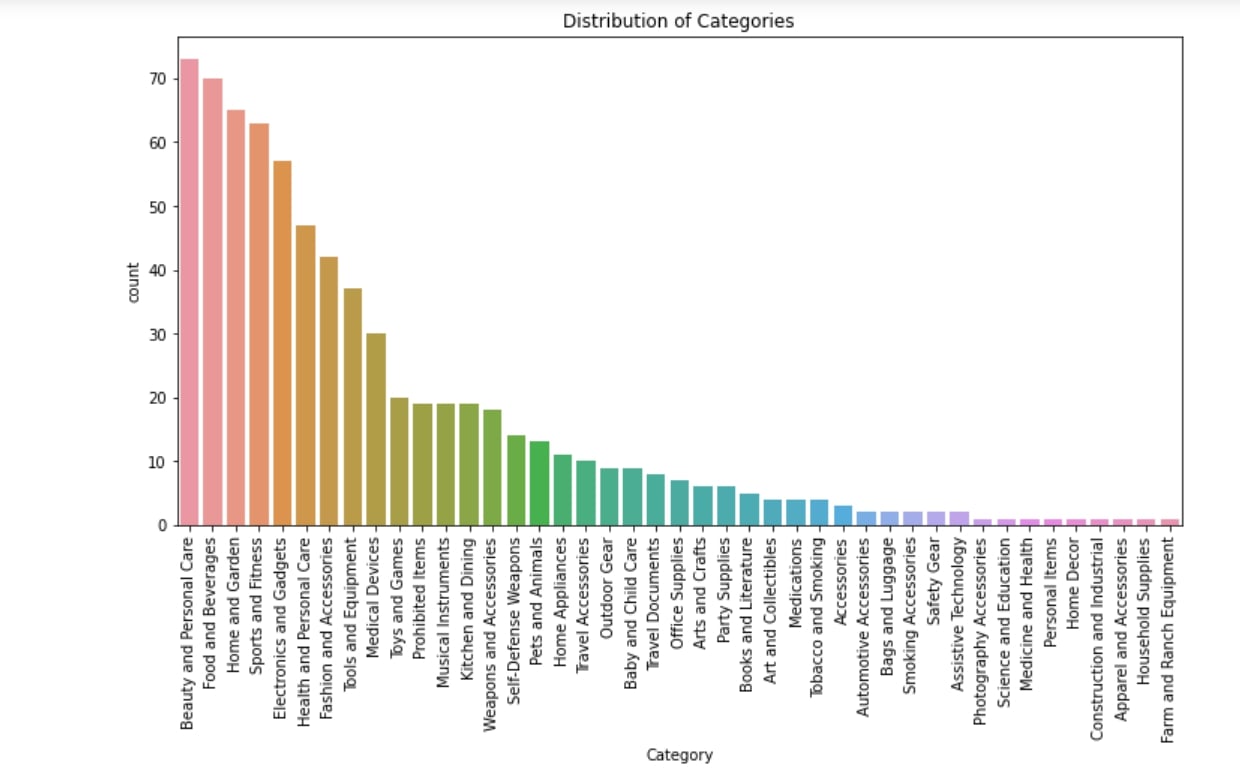}} \caption{Distribution of Categories items}
\label{fig:D}
\end{figure} 
\begin{figure}[H]
\centering
\fbox{\includegraphics[width=0.9\linewidth]{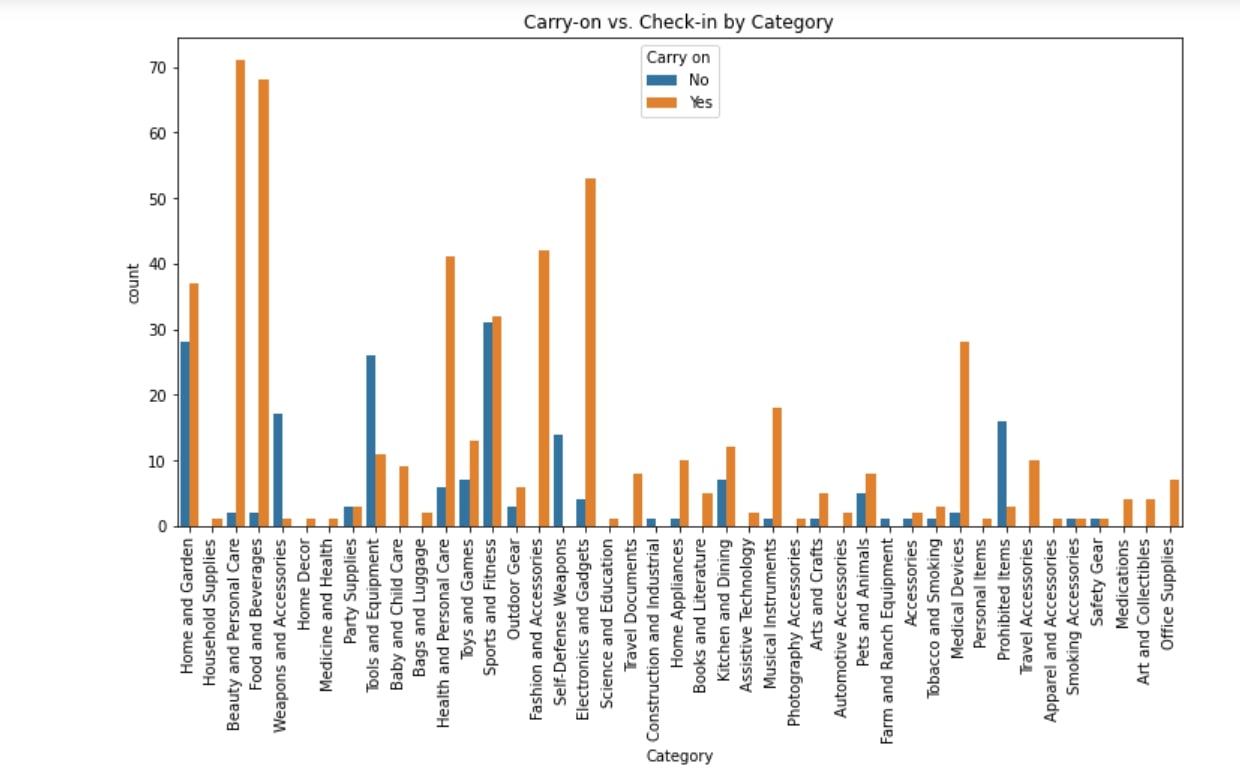}} \caption{Distribution of Carry on vs Check in items}
\label{fig:E}
\end{figure} 

\begin{figure}[H]
\centering
\fbox{\includegraphics[width=0.9\linewidth]{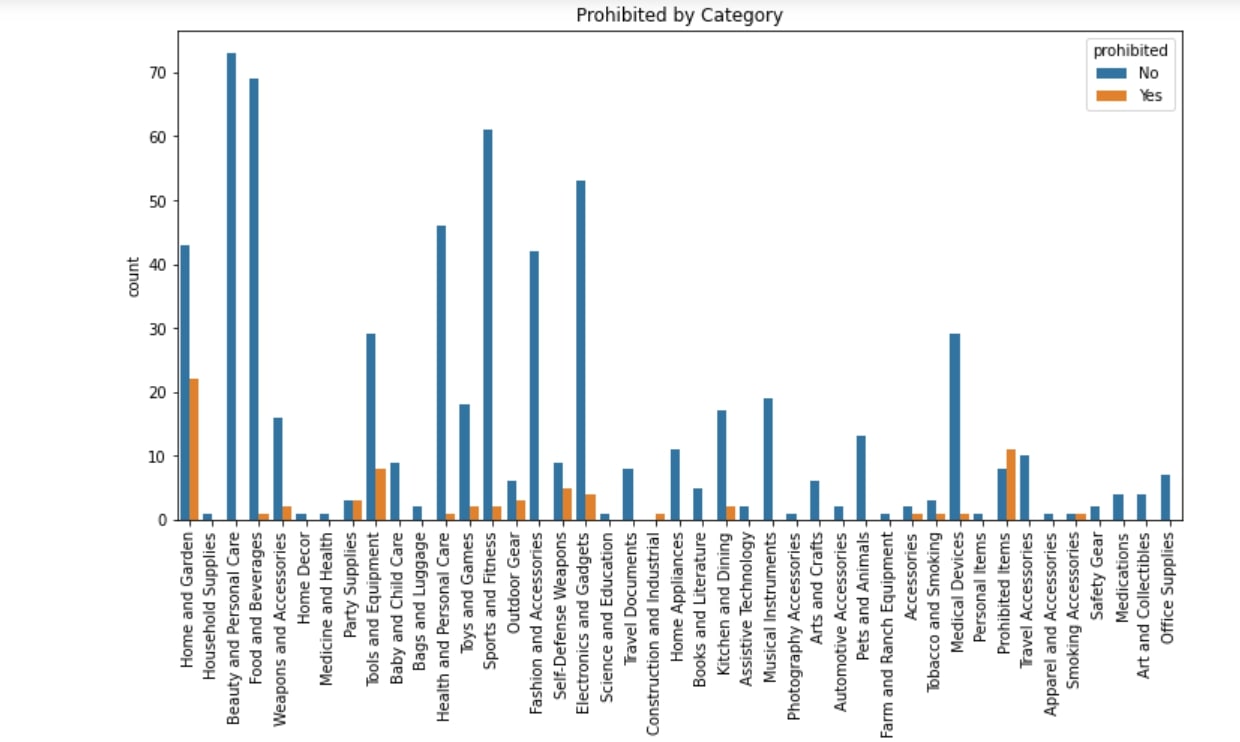}} \caption{Distribution of Prohibited by categories}
\label{fig:F}
\end{figure} 
Based on the dataset we explored the data visualization task. Shown in the Figure \ref{fig:A} contains the distribution of the carry\_on items out of the 712 items it distributed into yes and no as subcategories. Shown in the Figure \ref{fig:B} contains the distribution of the check\_in items out of the 712 items it distributed into yes and no as subcategories. Is shown Figure \ref{fig:C} contains the distribution of the prohibited\_items items out of the 712 items distributed into yes and no as subcategories. Shown in Figure \ref{fig:D} contains the distribution of the Categories items out of the 712 items it distributed 43 categories. Shown in Figure \ref{fig:E} contains the distribution of the Carry\_on vs Check\_in items out of the 712 items it distributed into subcategories like yes and no class. It shown in the Figure \ref{fig:F} contains the distribution of the prohibited by category out of the 712 items it distributed into subcategories like yes and no classes. Based on the visualization we are following the Proposed Frame work.
\section{Proposed Framework}
\begin{figure}[!ht]
\centering
\fbox{\includegraphics[width=0.9\linewidth]{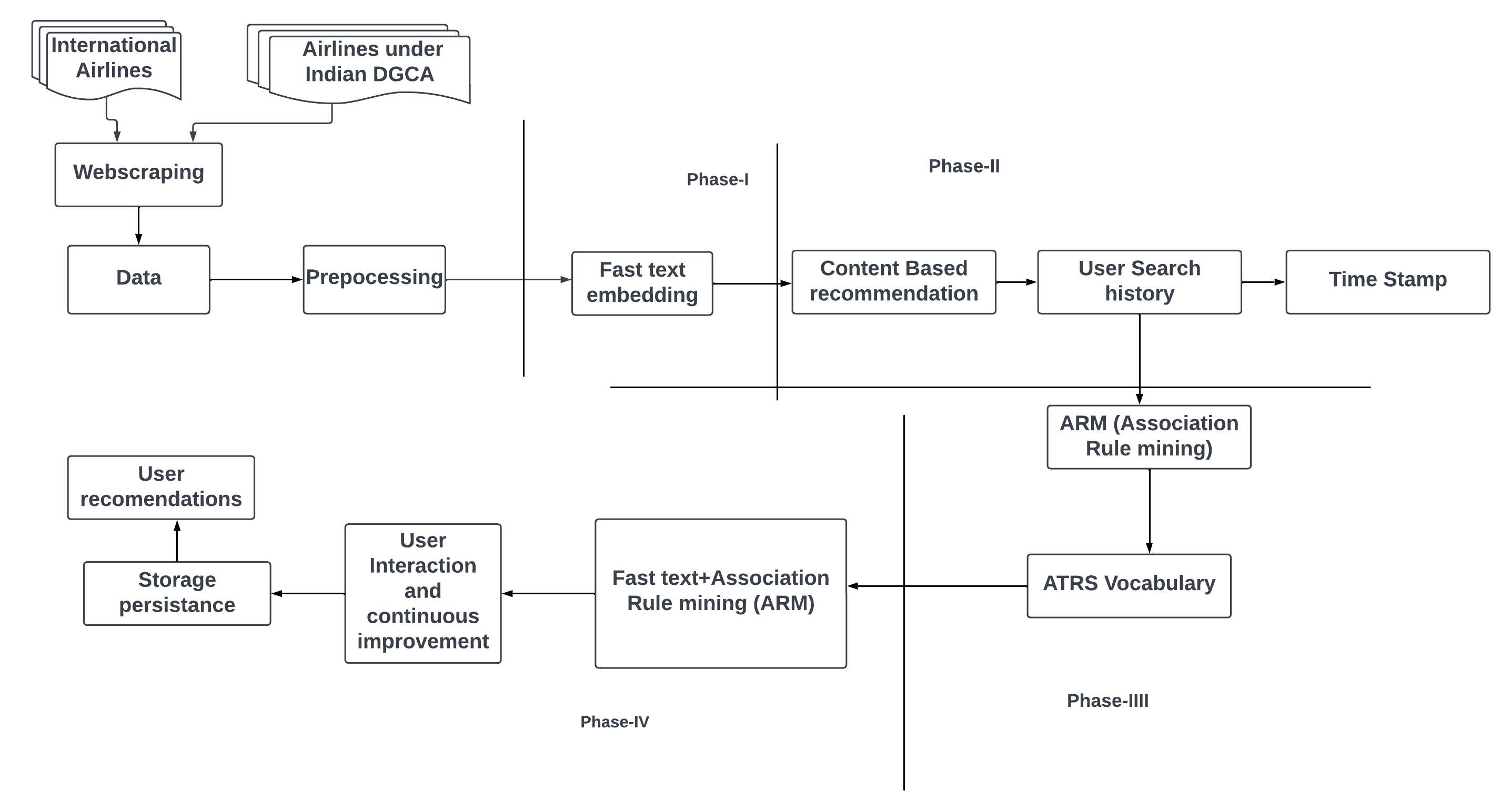}} \caption{ATRS General framework}
\label{fig:G}
\end{figure}
\begin{figure}
\centering
\fbox{\includegraphics[width=0.5\linewidth]{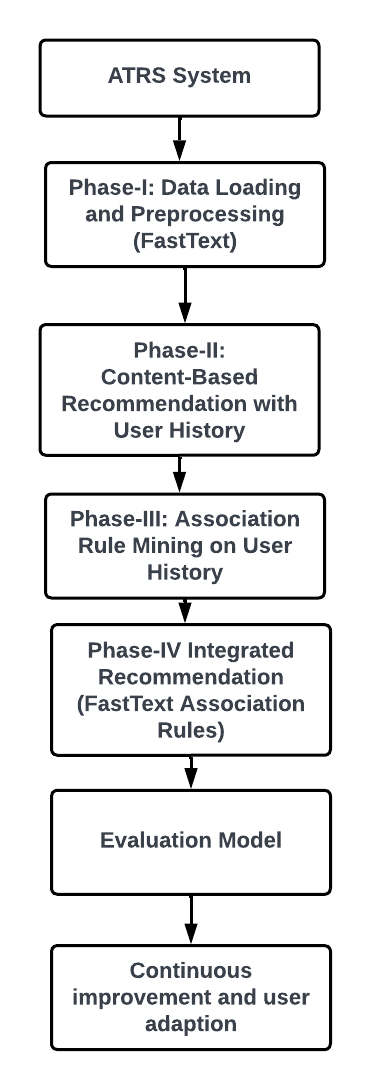}} \caption{Proposed framework}
\label{fig:H}
\end{figure}

In the proposed framework, Figure \ref{fig:G} provides a general overview of ATRS. Figure \ref{fig:H} illustrates the proposed framework of ATRS, functioning as a pre-travel advisory system for airports. It addresses queries such as which items are restricted from travel baggage during check-in, what items are permissible for carry-on baggage, and which items should be handed over to cabin staff after boarding the airplane. Below, we describe the proposed framework of ATRS, which initially targets international and domestic airlines under the Indian Directorate General of Civil Aviation (DGCA). We utilize web scraping techniques on airline websites to create the dataset outlined in the general framework, implementing each step according to our model. The proposed framework is an integral part of the general framework, detailing the phases of our model. 
\subsection{Web Scraping}  Web Scraping is a technique to extract data from one or more websites in the WWW (World Wide Web)\cite{zhao2017web}. And process it into simple structures such as spreadsheets, databases, or CSV files\cite{diouf2019web}. Commonly web data is scrapped utilizing Hyper Text Transfer Protocol (HTTP) through a web browser. It is a very complicated and time-consuming process. This is accomplished either manually by a user or automatically by a bot or web crawler (Finding and discovering URLs or Links). After exploring the above mentioned websites, we are using selenium web-driver, beautiful soup tools to scrape the data. After the Web Scraping and some URLs are blocked scraping process we choose the manual process and we build the dataset according to our requirements. It is stored in csv format.
\subsection{Data Preprocessing} After the CSV form of the data we apply tokenization on the items list. Here we applied the removal of the punctuation's, remove the space between the item name. In our case stemming, lemmatization, removal of stopwords will not apply on the dataset. After that, we build the clean dataset in the required format items name, carry-on, check-in, and prohibited items list and category. We are creating the new feature ItemDescription.  But lack of availability of the dataset we remove that feature. And only consider the remaining features. Which is stored in the CSV form of ATRS dataset.
\subsection{Fast-Text Embedding or Phase-I} Once the ATRS dataset build we load the fast text embedding (pre-trained vectors)\cite{athiwaratkun2018probabilistic}. The user input is tokenized, and FastText word embeddings are used to calculate the mean word embeddings for the tokens in the user input. Cosine similarity scores between the user input embeddings and the embeddings of all valid items are calculated. The top-N similar items are identified based on the similarity scores, and relevant information about the items is displayed to the user.And we categorize the data into 43 categories 
Figure \ref{fig:D} describe it. And we categorize the data into carry\_on items, check\_in items Figure \ref{fig:E} describe it. 
\begin{algorithm}
\caption{Phase-I Item Recommendation System}\label{alg:recommendation1}
\begin{algorithmic}[1]
\REQUIRE Data with columns: Item, Carry on, Check in, prohibited, Category
\STATE Import required libraries
\STATE Load CSV data and perform tokenization
\STATE Load FastText pre-trained word embeddings
\STATE Get unique categories and their count
\STATE Remove whitespace from column names
\STATE Combine columns to create ItemDescription
\STATE Define function \textsc{get\_recommendations(user\_input, top\_n)}
\WHILE{user input $\neq$ 'exit'}
    \STATE Get user input
    \STATE Get recommendations based on user input
    \STATE Filter data for exact match of user input
    \IF{user input matches an item}
        \STATE Display result for the entered item
    \ENDIF
    \STATE Display top similar items
\ENDWHILE
\end{algorithmic}
\end{algorithm}
here Phase-I results are displayed.
\begin{table*}[ht]
\caption{Our proposed model Phase-I results}
\label{table:ATRS1}
\centering
\begin{tabular}{|c|c|c|c|c|c|c|c|c|}
\hline
 \textbf{S.no}& \textbf{Item name}&\textbf{Carry on} &\textbf{Check in} &\textbf{Prohibited}&\textbf{\makecell{Top Similar\\ items Category}}\\
 \hline
\textbf{0}&aerosol paint & yes &yes&no &Aerosol\\
 \hline
 \textbf{1}&Gel Ice packs & yes &yes&no &Aerosol\\
 \hline
 \textbf{2}&Tear gas& no&no &yes &Aerosol\\
 \hline
 \textbf{3}&Magazine& yes&yes&no &Book \\
 \hline
 \textbf{4}&Comic books& yes&yes&no &Book \\
 \hline
 \textbf{5}&Newspaper& yes&yes&no &Book \\
 \hline
 \textbf{6}&Ipod& yes&yes&no & Laptop\\
 \hline
 \textbf{7}&DVD Players& yes&yes&no & Laptop\\
 \hline
 \textbf{8}&Desktop& yes&yes&no & Laptop\\
 \hline
 \textbf{9}&Power bank& yes&no&no &Electronics \\
 \hline
 \textbf{10}&Fuel cells& yes&yes&no &Electronics \\
 \hline
  \end{tabular}
\end{table*}
In the next step follows Phase-II recommendations for ATRS as follows:
\subsection{Phase-II: Recommendation using FastText and content based recommendations}
In the Phase-II ATRS follows that if any whitespace occurs in the column name we removed it. In our case there is no need for Lemmatization, stemming in our approach because we are only targeting the items which are pre-trained embeddings. We create the ItemDescription is the feature in the dataset due to lack of availability of the description of items we are skip the feature. And we apply tokenization on items. We apply the content based recommender systems (CBRS) on the user input items it will suggest the top similar items. These items are store in the user search history (Vocabulary) with time stamp. And it store into user search history. 
\begin{algorithm}
\caption{Phase-II Item Recommendation System Using (fast text + Content based Recommendations}\label{alg:recommendation2}
\begin{algorithmic}[2]
\REQUIRE Data with columns: Item, Carry on, Check in, prohibited, Category
\STATE Import required libraries
\STATE Remove whitespace from column names
\STATE Combine columns to create ItemDescription
\STATE Tokenize user input and calculate embeddings
\STATE Handle missing values and calculate similarity scores
\STATE Get top similar items based on similarity scores
\STATE Continuous user interaction loop:
\WHILE{user input $\neq$ 'exit'}
    \STATE Get user input
    \STATE Get content based recommendations based on user input
    \STATE Display result for entered item (if matched)
    \STATE Display top similar items (if available)
    \STATE Update user searches dictionary with timestamp
\ENDWHILE
\STATE Load existing user searches (if available)
\STATE Save user searches to CSV file
\end{algorithmic}
\end{algorithm}
Table \ref{table:ATRS2} describe the corresponding results.
\begin{table}[ht]
\caption{Our proposed model Phase-II results}
\label{table:ATRS2}
\begin{tabular}{lllllll}
\hline
\multicolumn{1}{|l|}{\textbf{S.no}} & \multicolumn{1}{l|}{\textbf{\begin{tabular}[c]{@{}l@{}}Item \\ name\end{tabular}}} & \multicolumn{1}{l|}{\textbf{\begin{tabular}[c]{@{}l@{}}Carry \\ on\end{tabular}}} & \multicolumn{1}{l|}{\textbf{\begin{tabular}[c]{@{}l@{}}Check \\ in\end{tabular}}} & \multicolumn{1}{l|}{\textbf{Prohibited}} & \multicolumn{1}{l|}{\textbf{\begin{tabular}[c]{@{}l@{}}Top\\ Similar\\ Items\\ Category\end{tabular}}} & \multicolumn{1}{l|}{\textbf{\begin{tabular}[c]{@{}l@{}}User\\ Search\\ History\end{tabular}}} \\ \hline
\multicolumn{1}{|l|}{0}             & \multicolumn{1}{l|}{Aerosol paint}                                                 & \multicolumn{1}{l|}{yes}                                                          & \multicolumn{1}{l|}{yes}                                                          & \multicolumn{1}{l|}{no}                  & \multicolumn{1}{l|}{aerosol}                                                                           & \multicolumn{1}{l|}{Aerosol paint}                                                            \\ \hline
\multicolumn{1}{|l|}{1}             & \multicolumn{1}{l|}{coffee}                                                        & \multicolumn{1}{l|}{yes}                                                          & \multicolumn{1}{l|}{yes}                                                          & \multicolumn{1}{l|}{no}                  & \multicolumn{1}{l|}{beverage}                                                                          & \multicolumn{1}{l|}{coffee}                                                                   \\ \hline
\multicolumn{1}{|l|}{2}             & \multicolumn{1}{l|}{ipod}                                                          & \multicolumn{1}{l|}{yes}                                                          & \multicolumn{1}{l|}{yes}                                                          & \multicolumn{1}{l|}{no}                  & \multicolumn{1}{l|}{laptop}                                                                            & \multicolumn{1}{l|}{ipod}                                                                     \\ \hline
\multicolumn{1}{|l|}{3}             & \multicolumn{1}{l|}{piano}                                                         & \multicolumn{1}{l|}{yes}                                                          & \multicolumn{1}{l|}{yes}                                                          & \multicolumn{1}{l|}{no}                  & \multicolumn{1}{l|}{instrument}                                                                        & \multicolumn{1}{l|}{piano}                                                                    \\ \hline             
\end{tabular}
\end{table}
In the next step Phase-III recommendations as follows:\\
\subsection{Phase-III ATRS (ARM using User Search History) and Time Stamp}
In the Phase-III ATRS follows that load the user searches data from csv file. Drop the index column from the dataset. Convert user searches data into transactional format, get unique items across all transactions. To create each transactional data frame into one hot encoding. Then we apply Apriori algorithm for finding frequent itemsets, to generate the association rules. Based on the user input and association rules it generate top\_n recommendations and there corresponding time stamp. And it store into the dictionary. If the user input is doesn't matched with in data then it will not recommended the item. If the user item fully matched or partially matched, or new item comes first time then it will append into user search history (dictionary). And getting top recommended items based on user input and association rules. If the recommended items exist in the dictionary it will display top recommended items otherwise it will not recommend.
\begin{algorithm}
\caption{Phase-III Association Rule Mining for Item Recommendations}\label{alg:association_rule_mining}
\begin{algorithmic}[3]
\REQUIRE CSV data with columns: Item, Carry on, Check in, Description
\STATE Load CSV data into a DataFrame
\STATE Load user searches data from CSV
\STATE Drop the index column from user searches DataFrame
\STATE Convert user searches data to transactional format
\STATE Get unique items across all transactions
\STATE Create DataFrame with one-hot encoding for each transaction
\STATE Apply Apriori algorithm to find frequent itemsets
\STATE Generate association rules
\STATE Define function \textsc{get\_recommendations(user\_input, association\_rules, top\_n)}
\STATE Get current timestamp
\STATE Create dictionary to store user searches with timestamp as key
\WHILE{true}
    \STATE Get user input
    \IF{user input is ``exit"}
        \STATE Exit loop
    \ENDIF
    \STATE Search for full matches of user input in data
    \IF{full match found}
        \STATE Display result for the entered item
        \STATE Append user search to dictionary
    \ELSE
        \STATE Search for partial matches of user input in data
        \IF{partial matches found}
            \STATE Display partial matches
            \STATE Append user search to dictionary
        \ENDIF
    \ENDIF
    \STATE Get top recommended items based on user input and association rules
    \IF{recommended items exist}
        \STATE Display top recommended items
    \ENDIF
\ENDWHILE
\STATE Convert dictionary to DataFrame
\STATE Load existing user searches data (if exists) and update with new searches
\STATE Save updated user searches to CSV file
\end{algorithmic}
\end{algorithm}
Here we are describing the Phase-III results.
\begin{table}[ht]
\caption{Our proposed model Phase-III results}
\label{table:ATRS3}
\begin{tabular}{|l|l|l|l|l|l|l|l|}
\hline
\textbf{S.No} & \textbf{\begin{tabular}[c]{@{}l@{}}Item \\ name\end{tabular}} & \textbf{\begin{tabular}[c]{@{}l@{}}Carry \\ on\end{tabular}} & \textbf{\begin{tabular}[c]{@{}l@{}}Check \\ in\end{tabular}} & \textbf{Prohibited} & \textbf{\begin{tabular}[c]{@{}l@{}}Top \\ similar\\ items\\  category\end{tabular}} & \textbf{\begin{tabular}[c]{@{}l@{}}User\\  search\\ history\end{tabular}} & \textbf{\begin{tabular}[c]{@{}l@{}}Time\\ stamp\end{tabular}} \\ \hline
0             & aerosol                                                       & yes                                                          & yes                                                          & no                  & aerosol                                                                             & nan                                                                       & 2023-07-29 19:26:07                                           \\ \hline
1             & coffee                                                        & yes                                                          & yes                                                          & no                  & beverage                                                                            & coffee                                                                    & 2023-07-31 12:51:50                                           \\ \hline
2             & ipod                                                          & yes                                                          & yes                                                          & no                  & laptop                                                                              & ipod                                                                      & 2023-07-29 19:35:44                                           \\ \hline
3             & piano                                                         & yes                                                          & yes                                                          & no                  & instruments                                                                         & piano                                                                     & 2023-07-31 13:00:39                                           \\ \hline
\end{tabular}
\end{table}
Here Phase-IV recommendations as follows:
\subsection{Phase-IV: ATRS using FastText and ARM}
\par{The algorithm utilizes Python libraries and techniques to process user search data from a CSV file, generate association rules using Apriori, and provide item recommendations based on user input and FastText embeddings. First, the algorithm loads user search data from a specified CSV file, excluding the index column, and converts it into transactions where each transaction contains non-null items from the DataFrame. Unique items across all transactions are extracted for one-hot encoding. Next, a one-hot encoded DataFrame is created with rows representing transactions and columns representing unique items. The Apriori algorithm is applied to discover frequent itemsets based on a specified minimum support threshold. Association rules are generated from these frequent itemsets using a minimum confidence threshold. The algorithm utilizes Python libraries and techniques to process user search data from a CSV file, generate association rules using Apriori, and provide item recommendations based on user input and FastText embeddings. First, the algorithm loads user search data from a specified CSV file, excluding the index column, and converts it into transactions where each transaction contains non-null items from the DataFrame. Unique items across all transactions are extracted for one-hot encoding.} 
Phase-IV algorithm explains Fast Text and ARM rules to recommends items to the user. Next, a one-hot encoded DataFrame is created with rows representing transactions and columns representing unique items. The Apriori algorithm is applied to discover frequent itemsets based on a specified minimum support threshold. Association rules are generated from these frequent itemsets using a minimum confidence threshold. A function processes user input to provide recommendations using FastText embeddings and association rules. It tokenizes the user input, calculates word embeddings, computes cosine similarity scores with item embeddings, and retrieves top-n similar items as recommendations. If valid embeddings are not available or contain NaN values, an empty list is returned. The algorithm combines data preprocessing, frequent itemset mining, association rule generation, and recommendation generation using embeddings. It demonstrates a comprehensive approach to leveraging data mining and natural language processing for personalized item recommendations based on user search.
\subsection{Evaluation of the Recommendation Model}
We evaluate the our recommendation model's performance using various metrics, including coverage, support, confidence, lift, leverage, and conviction. The evaluation metrics are calculated, visualized using a bar plot, and printed to assess the quality and effectiveness of the association rules in generating relevant recommendations for users.
\subsubsection{Improved Evaluation Metrics} 
In the context of association rule mining, support, confidence, lift, and leverage are commonly used metrics to assess the strength and significance of relationships between items in a dataset.
\subsubsection{Support} Support measures the proportion of transactions in the dataset that contain a particular itemset.\\
\textbf{Formula:} \( \text{Support}(X) = \frac{\text{Transactions containing } X}{\text{Total transactions}} \)
\subsubsection{Confidence} Confidence measures the likelihood that an item Y is purchased when item X is purchased.\\
 \textbf{Formula:} \( \text{Confidence}(X \rightarrow Y) = \frac{\text{Support}(X \cup Y)}{\text{Support}(X)} \)
\subsubsection{Lift} Lift measures how much more likely item Y is purchased when item X is purchased, compared to when Y is purchased without X.\\
\textbf{Formula:} \( \text{Lift}(X \rightarrow Y) = \frac{\text{Confidence}(X \rightarrow Y)}{\text{Support}(Y)} \)
\subsubsection{Leverage} Leverage measures the difference between the observed frequency of co-occurrence and the frequency that would be expected if X and Y were independent.\\
\textbf{Formula:} \( \text{Leverage}(X \rightarrow Y) = \text{Support}(X \cup Y) - (\text{Support}(X) \times \text{Support}(Y)) \)
In these formulas \(X\) and \(Y\) are itemsets or items. \(X \cup Y\) represents the union of sets X and Y. These metrics are commonly used in association rule mining to identify interesting and meaningful patterns in datasets, such as those that might be encountered in a baggage handling context. High support, confidence, lift, or leverage values indicate stronger associations between items.
\subsubsection{User Interaction and Continuous Improvement} The recommendation engine interacts with the user in a loop, accepting item names as input. It provides real-time recommendations based on the input and displays relevant item attributes. Additionally, the engine keeps track of user search history to personalize and improve future recommendations.
\subsubsection{Storage and Persistence} The engine stores user search history in a dictionary with timestamps and converts it into a DataFrame. This DataFrame is saved to the user\_searches.csv file for persistence and to maintain a history of user interactions.
The following section describe the results of the our proposed model.
\subsection{Comparative Analysis}
Comparative analysis between the ATRS dataset (Air Travel Recommender System dataset) and a market basket dataset involves examining the characteristics, goals, and applications of these datasets within their respective domains. Here's a detailed comparative analysis between the two datasets:
\begin{table}[htbp]
    \centering
    \caption{Comparative Analysis: ATRS vs Market Basket Dataset}
    \label{tab:comparative_analysis}
    \begin{tabular}{@{}|l|l|l|@{}}
    \hline
    \textbf{Aspect} & \textbf{ATRS Dataset} & \textbf{Market Basket Dataset} \\ 
    \hline
    Domain & Air travel recommendation & Retail and e-commerce \\
       \hline
    Focus & Personalized travel recommendations & Purchase behavior analysis \\
       \hline
    Data Characteristics & User profiles, transactional data & Item transactions, purchase history \\
       \hline
    Objective & Enhance travel planning & Improve product placement and sales \\
       \hline
    Analytical Techniques & Association rule mining, NLP & Apriori algorithm, Collaborative filtering \\
       \hline
    Challenges & Handling large-scale transaction data & Analyzing sparse and noisy data \\
       \hline
    Business Impact & Increased booking conversions & Optimized inventory management \\
       \hline
    \end{tabular}
\end{table}
In our proposed work the ATRS vocabulary data is works better results compare to the Market basket data. It shown in the Table \ref{table:ATRS5}. Figure \ref{fig:2} describe the evaluation metrics between ATRS (User\_searches.csv) and market basket data (store\_data.csv).
\section{Analysis of Experimental results}
The proposed intelligent baggage item recommendation system was evaluated using a comprehensive dataset of travel items and user interactions. The system incorporated four different recommendation approaches, to tackle the packing problems namely Phase-I, Phase-II, Phase-III, and Phase-IV, each utilizing different methodologies to provide item suggestions to users. The results of each approach are presented below:
\subsection{Phase-I Results analysis}
In Phase-I of implementing the ATRS (Air travel recommender system), several key steps are executed after the creation of the ATRS vocabulary. The size of initial vocabulary is 712 items.  We load the pre-trained FastText word embeddings \cite{athiwaratkun2018probabilistic}. The user input is tokenized, and FastText embeddings are used to compute the mean embeddings for the tokens in the user input. We then calculate cosine similarity scores between the user input embeddings and the embeddings of all valid items. Based on these similarity scores, the top-N similar items are identified, and relevant information about these items is displayed to the user. Additionally, the dataset is categorized into 43 distinct categories, which include categories for carry-on items and check-in items. Figure \ref{fig:D} provides a detailed description of this categorization process. Furthermore, the data is segmented into specific categories for carry-on and check-in items, as illustrated in Figure \ref{fig:E}. Table \ref{table:ATRS1}
describe the phase-I results. In the following step describe the Phase-II analysis.
\subsection{Phase-II Results analysis}
In Phase-II of implementing the ATRS (Air Travel Recommender System), several key steps are executed to prepare the dataset and facilitate recommendation tasks. Firstly, any whitespace in the column names of the dataset is removed to ensure uniformity and ease of data handling. Since our focus revolves around pre-trained embeddings of items, we do not employ lemmatization or stemming techniques for text normalization. The ``ItemDescription" feature is included in the dataset, but due to the absence of item descriptions, this feature remains empty or is omitted to avoid incomplete data. Each item undergoes tokenization to break them down into individual tokens for further processing. The ATRS utilizes content-based recommender systems (CBRS) to analyze item attributes and suggest top similar items based on content similarity \cite{zukerman2001predictive}. Recommended items are stored in the user search history, forming a vocabulary of user interactions with timestamps to track preferences and deliver personalized recommendations over time. This comprehensive approach ensures efficient data preparation and effective recommendation generation within the ATRS framework. Table \ref{table:ATRS2}
describe the phase-II results. In the following step describe the Phase-III analysis.
\subsection{Phase-III Results analysis}
In Phase-III of the ATRS (Air Travel Recommender System) implementation, the system processes user search data retrieved from a CSV file. Initially, the index column is dropped from the dataset to streamline subsequent operations. The user search data is then transformed into a transactional format, where unique items across all transactions are identified and organized. Each transactional dataset is encoded using one-hot encoding to represent item occurrences effectively. Next, the Apriori algorithm is applied to identify frequent itemsets and generate association rules based on the transactional data. These association rules form the basis for recommending items to users. When a user input is received, the ATRS leverages these association rules to generate top-N recommendations, along with corresponding timestamps. The system maintains a dictionary to store user interactions and recommendations. If a user input matches an item in the dataset, the corresponding recommendations are appended to the dictionary with timestamps. Conversely, if the user input does not match any data, no recommendations are provided. For matched user items, whether fully or partially, and for new items encountered for the first time, the system updates the user search history in the dictionary. The top recommended items based on the user input and association rules are displayed to the user. Recommendations are only displayed if they exist in the dictionary, ensuring that relevant and previously recommended items are accessible to users. This comprehensive process ensures efficient recommendation generation and user interaction tracking within the ATRS framework. Table \ref{table:ATRS3} describe the phase-III results. In the following step describe the Phase-IV analysis.
\subsection{Phase-IV Results analysis}
Phase-IV of the Air Travel Recommender System (ATRS) involves utilizing Python libraries and techniques to process user search data from a CSV file, generate association rules using the Apriori algorithm, and provide item recommendations based on user input and FastText embeddings. Initially, the algorithm loads user search data from a specified CSV file, excluding the index column, and converts it into transactions containing non-null items. It then extracts unique items from these transactions for one-hot encoding. A one-hot encoded DataFrame is created, with transactions as rows and unique items as columns. Next, the Apriori algorithm is applied to discover frequent itemsets based on a specified minimum support threshold. Association rules are generated from these frequent itemsets using a minimum confidence threshold. To provide recommendations, a function processes user input by tokenizing it, calculating word embeddings using FastText, and computing cosine similarity scores with item embeddings derived from the association rules. The function retrieves top-N similar items as recommendations based on these scores. If valid embeddings are unavailable or contain NaN values, an empty list is returned. Overall, Phase-IV combines data preprocessing, frequent itemset mining, association rule generation, and recommendation generation using embeddings. It demonstrates a comprehensive approach to leveraging data mining and natural language processing techniques for personalized item recommendations based on user search input within the context of air travel domain. Table \ref{table:ATRS4} describe the phase-IV results. In the following step describe the why it is required in ATRS.
\begin{table}[ht]
\caption{Our proposed model results}
\label{table:ATRS4}
\begin{tabular}{|l|l|l|l|l|l|}
\hline
\textbf{UID} & \textbf{Itemset recommendation}                & \textbf{support} & \textbf{confidence} & \textbf{lift} & \textbf{leverage} \\ \hline
0            & \{Piano\}                                      & 0.5              & 1.0                 & 2.0           & 0.25              \\ \hline
1            & \{Ipod \}                                      & 0.5              & 1.0                 & 2.0           & 0.25              \\ \hline
2            & \{Coffee\}                                     & 0.5              & 1.0                 & 2.0           & 0.25              \\ \hline
3            & \{piano, Ipod\}                                & 0.5              & 1.0                 & 2.0           & 0.25              \\ \hline
4            & \{Piano, Coffee\}                              & 0.5              & 1.0                 & 2.0           & 0.25              \\ \hline
5            & \{Ipod, Coffee\}                               & 0.5              & 1.0                 & 2.0           & 0.25              \\ \hline
6            & \{Ipod, Coffee, Piano\}                        & 0.5              & 1.0                 & 2.0           & 0.25              \\ \hline
..           & \{...\}                                        & ..               &                     &               &                   \\ \hline
109877       & \{baby powder, guitar, pickle, baby wipes,..\} & 0.25             & 1.0                 & 4.0           & 0.1875            \\ \hline
109878       & \{baby powder, flute, guitar, pickle, pi...\}  & 0.25             & 1.0                 & 4.0           & 0.1875            \\ \hline
\end{tabular}
\end{table}
\subsection{Why ATRS is Required}
Initially, our approach focuses on using a small dataset specific to Air Travel Recommender Systems (ATRS), which includes details of carry-on items, check-in items, and prohibited items during the packing process. In our scenario, we do not have access to user preference data. Instead, we treat transaction IDs as representing users. For each user, if they select a single item, we can determine whether it should be packed in carry-on, check-in baggage, or identified as a prohibited item to avoid discomfort during travel. The various item combinations, ranging from a single item up to a combination of 712 items in our approach, represent different user scenarios. We have collected approximately 109,878 user IDs and their corresponding item set recommendations. Packing these items appropriately ensures smooth airport handling without issues. Passengers can proceed directly to the check-in counter, saving queuing and waiting time. Our model aims to reduce the time required for packing baggage, thus helping passengers avoid overweight baggage fees.
\begin{table}[ht]
    \centering
    \begin{tabular}{|c|c|}
        \hline
        \textbf{Item Combination} & \textbf{User Id} \\
        \hline
        $\{i1\}$ & UID1 \\
        $\{i1, i2\}$ &  UID2 \\
        $\{i1, i2, i3\}$ &  UID3 \\
        $\{i1, i3, i4\}$ &  UID4 \\
        $\{i2, i3, i5\}$ &  UID5 \\
        $\{i1, i2, i3, i4,...,i712\}$ &  UIDn \\
        \hline
    \end{tabular}
    \caption{Mapping of Item Combinations to User transaction IDs (UIDs)}
    \label{tab:item_tid_mapping}
\end{table}
 Each combination reflects a specific set of chosen items by the user, demonstrating the flexibility and scalability of our recommendation system to accommodate diverse user preferences and selections. In the future we work on the user preferences and selections we may identified the purpose of the traveler. If he/she is a tourist what kind of items he may be brings, if he/she may be musician what type of instruments he may be carried into the carry-on, check-in baggage. Like that we can identified the tourist behaviour. If the Item Descriptions is available we can use the Bert (sentence embeddings). 
\begin{table*}[ht]
\caption{On applying over the market basket dataset our model results}
\label{table:ATRS5}
\centering
\begin{tabular}{|c|c|c|c|c|c|c|c|c|}
\hline
 \textbf{UID}& \textbf{Itemset Recommendation}&\textbf{support} &\textbf{Confidence} &\textbf{Lift}&\textbf{leverage}\\
 \hline
\textbf{0}&\makecell{\{{eggs, ground beef}\}}&   0.010 &0.50 &2.12 &0.005\\
 \hline
 \textbf{1}&\makecell{\{{milk, ground beef}\}}&  0.011 &0.50& 2.11 &0.005\\
 \hline
  \textbf{1}&\makecell{\{{milk, ground beef,bread}\}}&  0.011 &0.50& 2.10 &0.005\\
 \hline
 \end{tabular}
\end{table*}
\begin{figure}[ht]
\centering
\fbox{\includegraphics[width=0.9\linewidth]{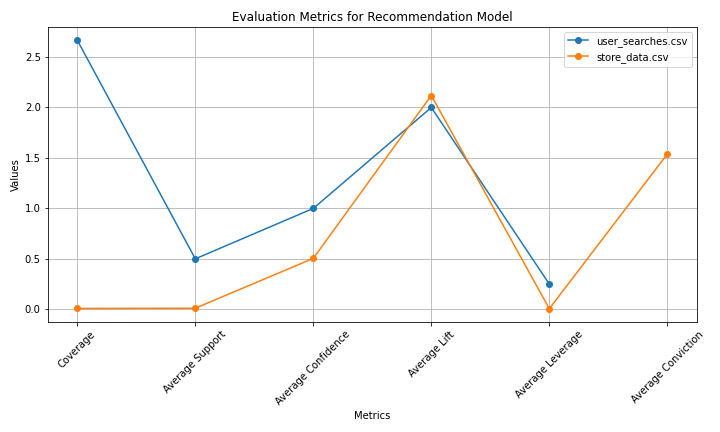}} \caption{Evaluation metrics}
\label{fig:2}
\end{figure} 
\subsection{Association Rules metrics}
The following steps describes the association rules metrics \cite{unvan2021market}.
\begin{itemize}
    \item \textbf{Support} Support quantifies the prevalence of a specific itemset within the dataset. It is calculated as the proportion of transactions containing the itemset.
   \item \textbf{Interpretation:} A higher support value indicates a more frequently occurring itemset.
   \item \textbf{Support} Support quantifies the prevalence of a specific itemset within the dataset. It is calculated as the proportion of transactions containing the itemset.
   \item \textbf{Interpretation:} A higher support value indicates a more frequently occurring itemset.
\item \textbf{Confidence} Confidence measures the likelihood that when item X is present, item Y is also present. It is computed as the ratio of the support for the combined itemset to the support for item X alone.
\item \textbf{Interpretation:} Higher confidence values signify stronger associations between items.
\item \textbf{Lift} Lift assesses the strength of the association between two items by comparing the observed probability of their co-occurrence to the expected probability if the items were independent.
\item \textbf{Interpretation:} A lift value greater than 1 indicates a positive association, suggesting that the presence of one item increases the likelihood of the other.
\item \textbf{Leverage} Leverage measures the difference between the observed co-occurrence of two items and the expected co-occurrence assuming independence.
\item \textbf{Interpretation:} Higher leverage values indicate a higher-than-expected frequency of co-occurrence, suggesting a non-random association.
\end{itemize}
In the realm of baggage handling, association rule metrics serve as invaluable tools for uncovering relationships and patterns within datasets. Specifically, a high confidence value in a rule pertaining to the presence of certain items in baggage signals a robust predictive relationship. Meanwhile, lift and leverage measurements aid in gauging practical significance and assessing deviation from independence, providing crucial insights for optimizing baggage handling processes. The evaluation of the recommendation system demonstrated its potential in revolutionizing the travel industry and other domains that require personalized item recommendations. By leveraging advanced techniques and historical user interactions, the system offered users valuable insights and assistance in making well-informed packing decisions. As technology continues to evolve and more user data becomes available, the system is expected to further improve and adapt to the dynamic needs of travelers, enhancing their travel experiences and overall satisfaction. We are explaining our model workforce one use\_search.csv data with one example it shown in the Table \ref{table:ATRS4}, Table \ref{tab:item_tid_mapping} describe  the user carried items like single item, or combination of multiple items to carry into the baggage, And comparing our model workforce with market\_basket (store\_data.csv) it shown in the Table \ref{table:ATRS5}. And the Figure \ref{fig:2} shows the metrics of our model. Based on the metrics our models are suggest the items to the user what are the items are carried into the baggage. In the user history  the combinations of the items are more our model give good results. The metrics results are shown in the figure \ref{fig:2}, in that figure blue color line indicates our proposed model it's work on user\_search.csv and orange color line indicates our model compared with market\_basket (store\_data.csv). If the use\_search history is more our model gives good results. The following section describes discussion and concluding remarks.
\subsection{Benefits of Air Travel Baggage Recommendation Systems}
Air travel baggage recommendation systems offer several benefits that enhance the overall travel experience for passengers. Some key benefits include:
\begin{enumerate}
    \item \textbf{Efficiency}: These systems help travelers optimize their packing by recommending the most suitable items based on destination, trip duration, weather conditions, and personal preferences. This leads to efficient use of luggage space and reduces the likelihood of overpacking or forgetting essential items.
    \item \textbf{Convenience}: By providing personalized recommendations, baggage systems simplify the packing process for travelers. This convenience saves time and effort, especially for frequent flyers who may have varying travel needs. 
    \item \textbf{Enhanced Travel Experience}: Tailored recommendations contribute to a smoother travel experience by ensuring that passengers are well-prepared for their journey. This can reduce stress associated with packing and increase overall satisfaction with the trip.
    \item \textbf{Improved Decision-Making}: Baggage recommendation systems leverage data analytics and user profiling to deliver informed suggestions. This helps travelers make better decisions about what to pack, considering factors they may not have thought of on their own.
    \item \textbf{Adaptability}: These systems can adapt to changing circumstances, such as weather updates, flight delays, or itinerary changes. By providing real-time recommendations, they enable travelers to adjust their packing accordingly.
    \item \textbf{Personalization}: Personalized recommendations cater to individual preferences and travel habits, enhancing the relevance and usefulness of the suggestions provided. This personal touch contributes to a more tailored and enjoyable travel experience.
    \item \textbf{Resource Optimization}: Efficient packing reduces excess baggage fees, minimizes environmental impact, and improves operational efficiency for airlines. By encouraging optimal packing practices, these systems support sustainability in air travel.
\end{enumerate}
Overall, air travel baggage recommendation systems leverage technology to enhance packing efficiency, simplify decision-making, and elevate the travel experience for passengers, ultimately contributing to smoother and more enjoyable journeys.
\section{Discussion and Concluding Remarks}
This section highlights the adaptability of the recommendation system, ethical considerations in data handling, and the integration of user feedback as key elements contributing to a more effective and user-centric baggage item recommendation system.
\subsection{Handling updates or adaptation to evolving of user preferences over time}
The system is designed to handle updates and adapt to evolving user preferences over time through a dynamic data management approach. As new items or user preferences are introduced, the system incorporates them into a centralized data repository, such as a dictionary or a CSV file. This repository serves as a comprehensive and up-to-date source of information, continuously expanding as the dataset grows. The model utilizes this evolving dataset to refine its understanding of user preferences and adapt its recommendations accordingly \cite{ricci2015user}. By regularly updating the dataset with new interactions and preferences, the system ensures that it stays current and responsive to changes in user behavior, ultimately enhancing the accuracy and relevance of its recommendations over time.
\subsection{Ethical considerations, user privacy measures and data security concerns}
In our data handling model, paramount attention is given to ethical considerations and user privacy. To safeguard sensitive information, we adhere to a strict data minimization approach. This involves the complete removal of personally identifiable information (PII) such as phone numbers and addresses from our datasets. Additionally, to preserve user anonymity while retaining the capacity to recognize patterns and preferences, we replace usernames with anonymous user IDs. This anonymization process not only protects user privacy but also ensures compliance with ethical data practices. We prioritize transparency by informing users about these privacy measures, obtaining their informed consent, and offering them control over their data. Our commitment extends to regular audits, robust security measures, and ongoing employee education to uphold the highest standards of ethical data handling and user privacy protection.
\subsection{Integrating user feedback and improvement of personalized recommendations}
In our ongoing commitment to enhancing user experience and maintaining a user-centric approach, we envision the integration of user feedback mechanisms as a pivotal aspect of our system's future development. Recognizing the dynamic nature of user preferences, this enhancement aims to establish a direct channel for users to provide feedback on recommendations. By incorporating user insights, preferences, and critiques, we anticipate a continuous improvement cycle, allowing the system to adapt and refine its algorithms \cite{ricci2015user}. This iterative process not only ensures that user feedback plays a central role in shaping the system's recommendations but also fosters a personalized experience tailored to individual preferences. Through this forward-looking initiative, we aim to cultivate a more responsive and user-friendly system, aligning with our commitment to delivering recommendations that genuinely resonate with user needs and expectations.
\subsection{Conclusion and Future scope}
In this research, we have presented a comprehensive framework for the development and deployment of the Air Travel Recommender System (ATRS), which leverages advanced data processing, content analysis, and recommendation techniques tailored to the air travel domain. Through a systematic breakdown of the implementation phases, we have highlighted the essential steps and methodologies employed to ensure the system's efficiency, relevance, and effectiveness in providing personalized recommendations to users. The initial phases of the ATRS implementation focus on data preprocessing, user input processing, and content analysis. Tokenization of user queries and computation of FastText embeddings enable us to understand user intent and identify relevant items based on semantic similarity. Furthermore, the categorization of items into distinct categories such as carry-on and check-in items enhances the recommendation process by narrowing down choices and improving relevance. As we progress into subsequent phases, including dataset preparation, association rule mining, and recommendation generation, the ATRS demonstrates its capability to adapt and learn from user interactions. By leveraging techniques like the Apriori algorithm for frequent itemset mining and association rule generation, the system establishes meaningful relationships between items and translates these into actionable recommendations based on user queries. One of the key strengths of the ATRS lies in its ability to track user interactions over time, maintaining a comprehensive history of user preferences and search patterns. This enables the system to deliver increasingly personalized recommendations and adapt its suggestions based on evolving user needs. In conclusion, the phased implementation of the Air Travel Recommender System exemplifies a holistic approach to recommendation systems within the air travel domain. By integrating data mining, natural language processing, and content-based recommendation techniques, the ATRS offers a robust solution for enhancing user experience and facilitating informed decision-making during air travel. Moving forward, further enhancements and refinements to the system can continue to optimize recommendation quality and user satisfaction in the dynamic and evolving landscape of air travel services.
\begin{algorithm}
    \SetAlgoLined
    \SetKwInOut{Input}{Input}
    \SetKwInOut{Output}{Output}
    \Input{CSV file  (\texttt{csv\_file})}
    \Output{Association rules (\texttt{association\_rules})} 
    \BlankLine
    \caption{Phase-IV Fast Text and ARM based recommendations}
    \label{algo:association_rules}
    \BlankLine
    \tcp{Load user searches data from CSV}
    \texttt{df} $\leftarrow$ Read CSV file (\texttt{csv\_file})\;
    \BlankLine
    \tcp{Drop the first column (index) which is not required}
    \texttt{df} $\leftarrow$ \texttt{df.iloc[:, 1:]}\;
    \BlankLine
    \tcp{Convert data to a transactional format}
    \texttt{transactions} $\leftarrow$ List of lists\;
    \For{\textnormal{each row} \texttt{in} \texttt{df}}{
        \texttt{transaction} $\leftarrow$ List of non-null items in \texttt{row}\;
        \texttt{transactions.append(transaction)}\;
    }
    \BlankLine
    \tcp{Get unique items across all transactions}
    \texttt{all\_items} $\leftarrow$ List of unique items in \texttt{transactions}\;
    \BlankLine
    \tcp{Create a DataFrame with one-hot encoding for each transaction}
    \texttt{transactions\_df} $\leftarrow$ Create one-hot encoded DataFrame for \texttt{transactions} using \texttt{all\_items}\;
    \BlankLine
    \tcp{Apply Apriori algorithm to find frequent itemsets}
    \texttt{min\_support} $\leftarrow$ 0.1 \tcp*{Adjust this threshold based on your data}
    \texttt{frequent\_itemsets} $\leftarrow$ Apply Apriori to \texttt{transactions\_df} with \texttt{min\_support}\;
    \BlankLine
    \tcp{Generate association rules}
    \texttt{min\_confidence} $\leftarrow$ 0.5 \tcp*{Adjust this threshold based on your data}
    \texttt{association\_rules} $\leftarrow$ Generate rules from \texttt{frequent\_itemsets} with \texttt{min\_confidence}\;
    \BlankLine
    \tcp{Function to get recommendations using FastText embeddings and ARM}
    \SetKwFunction{FMain}{get\_recommendations}
    \SetKwProg{Fn}{def}{:}{}
    \Fn{\FMain{\textnormal{user\_input}, \textnormal{association\_rules}, \textnormal{top\_n=5}}}{
        \tcp{Tokenize user input}
        \texttt{user\_input\_tokens} $\leftarrow$ Lowercased tokens of \texttt{user\_input}\;
        \tcp{Calculate word embeddings for user input}
        \texttt{user\_input\_embeddings} $\leftarrow$ Mean embeddings of \texttt{user\_input\_tokens}\;
        \tcp{Check for valid embeddings}
        \If{\texttt{user\_input\_embeddings} is not None and no NaN values}{
            \tcp{Compute similarity scores with item embeddings}
            \texttt{similarity\_scores} $\leftarrow$ Cosine similarity between \texttt{user\_input\_embeddings} and item embeddings\;
            \tcp{Get top-n similar items}
            \texttt{top\_indices} $\leftarrow$ Indices of top-n items based on \texttt{similarity\_scores}\;
            \tcp{Prepare recommendations}
            \texttt{recommendations} $\leftarrow$ List of recommended items based on \texttt{top\_indices}\;
            \KwRet \texttt{recommendations}\;
        }
        \Else{
            \KwRet Empty list\;
        }}
\end{algorithm}
\subsubsection{Acknowledgement}
Mudavath Ravi extend express his heartfelt gratitude to Prof. Sanjay Chitnis as he was instrumental in motivating and guiding the exploration of this research problem.
\section{Declaration}
Authors declare no conflict of interest. \\
\textbf{Ethical Approval:} Ethical approval was not required. \\
\textbf{Funding:} This research was conducted without external financial support. The authors declare that there was no specific funding received for this study.\\
\textbf{Availability of data and materials:}
The authors will make data available on request.\\
\bibliographystyle{splncs04}
\bibliography{ATRS}
\end{document}